# Accepted manuscript

Title： Efficiency and characteristics of MICP in environments with elevated salinity, diminished oxygen, and lowered temperature: a microfluidics investigation

# Efficiency and Characteristics of MICP in Environments with Elevated Salinity, Diminished Oxygen, and Lowered Temperature: A Microfluidics Investigation

Jianyu Yang[1] and Yuze Wang[2*]

**Abstract:**

Microbially Induced Carbonate Precipitation (MICP) shows significant potential for improving soil strength, but environmental factors greatly influence its mechanisms and effectiveness. The feasibility of using MICP in demanding conditions, such as the soil around piles in shallow seabeds during winter, known for their elevated salinity, diminished oxygen, and lowered temperature remains uncertain due to limited research in this area. To address this gap, microfluidic techniques and advanced measurement tools, including Raman spectroscopy and SEM, were utilized to investigate the impact of high salinity, low oxygen levels, and cold temperatures on bacterial growth, calcium carbonate crystallization, and porous medium permeability through MICP. The findings reveal that cold temperatures notably hinder bacterial growth, while high salinity and low oxygen levels also play significant roles. Low oxygen levels particularly reduce bacterial attachment. Additionally, in seawater environments, high salinity and cold temperatures have a more pronounced effect on calcium carbonate crystal shape and type, while the impact of low oxygen levels is relatively minor. Specifically, high salinity has minimal effect on average crystal diameter but reduces crystal number by 20.2%, while low oxygen levels increase average diameter (20.3%) but decrease crystal quantity (50.9%). Furthermore, cold temperatures decrease average diameter (36.9%) with little impact on crystal quantity. After six injections of the cementation solution, the chemical transformation efficiency of MICP-treated samples under combined marine conditions (high salinity, low oxygen levels, and cold temperatures) is 20.6% of DI water, atmospheric oxygen levels, and 20°C, with cold temperatures being the primary contributor to this reduction (40.1%). An exponential decline in permeability with increasing calcium carbonate content is also observed. Based on the location of calcium carbonate generation within the percolation channel, it can be categorized into two types: fast decay and slow decay. Overall, this study provides valuable insights for optimizing mineralization processes in challenging marine conditions and opens new avenues for stabilizing shallow seafloors using MICP. Additionally, this study highlights the 'induced' nature of MICP, illustrating how bacterial activity, along with environmental factors, impacts its performance, thus posing challenges for its application in engineering projects with varying environmental conditions.

1. Ph.D. Student, Dept. of Ocean Science and Engineering, Southern Univ. of Science and Technology, Shenzhen, CN 518055. E-mail: 12031074@mail.sustech.edu.cn
2. Associate Professor, Dept. of Ocean Science and Engineering, Southern Univ. of Science and Technology, Shenzhen, CN 518055 (Corresponding author). E-mail: wangyz@sustech.edu.cn

# Introduction

Preventing damage to seabed pipelines and cables due to marine landslides (Carter, 2009; Wang et al., 2018; Dutta et al., 2019), as well as the destabilization of foundations for offshore installations due to soil erosion (Chen et al., 2022), has become critically important for the sustainable exploitation of marine resources. Cement piles, implemented in areas susceptible to landslides and erosion, constitute the primary solution to these challenges (Lirer, 2012; Hu et al., 2021). However, the characteristics of the soil surrounding the pile significantly influence its overall performance. Microbially Induced Carbonate Precipitation (MICP) has been used in recent years to reinforce soil surrounding piles due to its ability to increase soil strength by coating or bridging particles with $CaCO_3$ (Whiffin, 2004; Wang et al., 2019a; Wang et al., 2019b). This method effectively improves the soil surrounding the pile and increases the bearing capacity of the pile foundation (Lin et al., 2016; Lin et al., 2018; Lin et al., 2021). These studies were conducted under the condition of deionized water at room temperature. Other research showed that under salinity condition, MICP improved the resistance of samples to erosion (Li et al., 2023a; Li et al., 2023b). However, the environmental factors impacting actual marine engineering are complex due to their variability and the multitude of interacting elements involved. For example, in the East China Sea near Fujian and Zhejiang Province, the temperature of seawater in winter is 7-17°C. In particular, the temperature at the mouth of the Yangtze River Delta hovers around 10°C in winter (Zheng et al., 1982; Wang et al., 2021a). The salinity ranges from 22 to 34 practical salinity units (PSU: grams of dissolved substance per kilogram of water) (Delcroix et al., 2002; Chen et al., 2006), and the dissolved oxygen (DO) level at a depth of 20 meters remains stable at 2-3 mg/L, constrained by limited oxygen diffusion from the water column (Chen et al., 2007). These factors significantly affect bacterial growth, urease activity and mineralization efficiency during MICP.

Currently, the bacterial strains used for MICP studies in marine environments are mainly *Sporosarcina pasteurii* (*S. pasteurii*), *Sporosarcina aquimarina* (*S. aquimarina*), *Sporosarcina newyorkensis* (*S. newyorkensis*), *Bacillus megaterium* (*B. megaterium*) and various urease bacteria isolated *in situ* (Jiang et al., 2016; Al Imran et al., 2019; Saracho et al., 2020; Ikoma et al., 2021; Lin et al., 2023). The tolerance of different strains mainly depends on the conditions of their isolation environment. *S. aquimarina*, isolated from Korean seawater, has good salinity tolerance (Yoon et al., 2001). *S. newyorkensis*, isolated for the first time from a hydrate reservoir, has a good tolerance to low temperatures (Hata et al., 2020). *B. megaterium* has good tolerance to both oxygen and low temperature (Jiang et al., 2016; Sun et al., 2019b). Among all urease bacteria, *S. pasteurii* has become a commonly used strain for MICP applications under marine conditions, owing to its high rate of catalytic urea hydrolysis (Terzis et al., 2019) and extreme environmental adaptability (Kannan et al., 2020). Field experiments using *S. pasteurii*, such as coastal erosion control (Li et al., 2024) and high-pressure, high-temperature oil well remediation (Phillips et al., 2018), have shown that *S. pasteurii* is highly efficient at producing calcium carbonate even under extreme conditions. Nevertheless, *S. pasteurii* is still subject to extreme conditions such as salinity, temperature and oxygen, and has tolerance limits to these extremes.

Salinity plays an important role in MICP. While *S. pasteurii* can thrive and exhibit urease activity in marine culture medium (Lin et al., 2023; Xiao et al., 2023), high salinity impedes its growth and urease activity (Dikshit et al., 2022; Fu et al., 2022; Zhao et al., 2023), leading to a decrease in the rate of urea hydrolysis and subsequent calcium carbonate precipitation. Cheng et al.(2014) mixed seawater (~35 psu) with a bacterial solution for durations of 0 and 24 hours, respectively, and observed that bacterial urease activity decreased to 94.6% and 97.1% of control activity for each duration (Cheng et al., 2014b). Dong et al. (2021) measured the 48-hour growth curves of *S. pasteurii* in both seawater medium (~35 psu) and deionized water medium. After 48 hours of incubation, the specific urease activity in the seawater medium decreased to 76.2% compared to that in deionized water (Dong et al., 2021). *S. pasteurii* ceased growing and lost most of its urease activity at 80 psu, demonstrating its limits under high salinity conditions (Zhao et al., 2023). Additionally, magnesium ions present in seawater interfere with the formation of calcite during MICP, influencing the type of calcium carbonate formed (Noyes, 1962; Davis et al., 2000). The total amount and crystal type of calcium carbonate significantly impact the strength of the soil after MICP cementation. Some studies have demonstrated successful MICP bio-cementation in high-salinity marine environments, achieving considerable compressive strength in sand columns (Mortensen et al., 2011; Cheng et al., 2014b). However, conflicting findings exist, with some research indicating elevated salinity enhanced calcium carbonate precipitation rates (Mortensen et al., 2011), while others observed lower calcium carbonate content and strength in samples formed with natural seawater compared to freshwater (Peng et al., 2022). The inhibitory effect of metal ions on bacterial growth and urease activity in seawater-bonded sand columns potentially contributes to these variations (Lin et al., 2023). Peng et al. (2022) observed slightly lower calcium carbonate content in freshwater MICP samples compared to seawater samples, but higher UCS and tensile splitting strength (TSS) in the freshwater environment (Peng et al., 2022).

Diminished oxygen plays a crucial role in the MICP process. Bergey's manual describes *S. pasteurii* as an aerobic bacteria that can slightly grow in sugar-free anaerobic media (Whitman et al., 2015). Under diminished oxygen condition, *S. pasteurii* can still grow, albeit at a slightly reduced rate (Jain et al., 2019). However, *S. pasteurii* cannot grow in strictly anaerobic environments without dissolved oxygen (Martin et al., 2012; Jain et al., 2019). While diminished oxygen has minimal impact on urease activity (Whiffin, 2004; Jiang et al., 2016; Lapierre et al., 2024), it affects the metabolic activity and growth rate of microorganisms, influencing the morphology and rate of calcium carbonate precipitation. In sand column experiments, Li et al. (2018b) discovered that under oxygen-restricted conditions, the UCS strength of MICP-bonded samples was significantly lower (less than 0.1 MPa) compared to normal aerobic conditions (0.21-0.39 MPa) or air-supply conditions (0.43-1.8 MPa). This finding highlight how diminished oxygen can hinder the efficiency of MICP (Li et al., 2018).

Temperature is the key parameter in regulating the cementation effect of MICP. According to Bergey's Manual, the growth temperature range for *S. pasteurii* spans from 4°C to 42°C (Whitman et al., 2015). Urease activity, however, displays a broader tolerance, remaining active between temperatures of 4°C and 80°C, allowing for potential utility across diverse thermal conditions (Whiffin, 2004; Wang et al., 2023c). Both urease activity and its rate of decline increase with rising temperatures. For instance, at 4°C, urease activity is only about 12.4% of its level at the optimal temperature of 30°C, yet this activity can be sustained for over 10 days. Conversely, at 50°C, while initial urease activity is roughly three times higher than at the optimal temperature, it rapidly diminishes, falling to nearly zero within 24 hours (Wang et al., 2023c). Kim et al. (2018) discovered that the maximum amount of calcium carbonate was precipitated at 30°C in a range from 20-50°C, based on solution experiments (Kim et al., 2018). Similarly, Sun et al. (2019) reported that the amount of calcium carbonate precipitated increased with temperature over a period of 48 hours at temperatures between 15-30°C (Sun et al., 2019a). Previous studies have revealed that lowered temperature inhibited bacterial growth and activity, impacting the rate of urea hydrolysis and the precipitation of $CaCO_3$ (Wang et al., 2023c). For instance, equations fitted to urease activity with temperature data by van Paassen et al. (2009) and Wang et al. (2023) showed that urease activity at 10°C (the mouth of the Yangtze River Delta in winter) was approximately 47.6% and 45.6% of urease activity at 20°C, respectively (Van Paassen, 2009; Wang et al., 2023c). Moreover, according to nucleation theory (Zhang et al., 2002; Tong et al., 2004; Yu et al., 2004), lowered temperature increases the activation energy for crystal nucleation, thereby limiting the precipitation rate of calcium carbonate. Wang et al. (2023) found that while low temperature (4°C) did not diminish urease activity, it restricted bacterial growth, attachment rate, and the rate of calcium carbonate precipitation (Wang et al., 2023c).

Despite individual studies on the effects of elevated salinity, diminished oxygen, and lowered temperature on MICP, there remains a notable gap in comprehensive research examining their combined influence. Consequently, the mechanism of MICP processes based on the main contributing factor remains unclear. To address this knowledge gap, this study investigates the collective impact of elevated salinity, diminished oxygen, and lowered temperature on MICP by utilizing innovative microfluidic technology to simulate MICP in shallow seafloors. We comprehensively analyze the effects on calcium carbonate size, type, growth process, and permeability. Through the examination of crystal characteristics using Raman spectroscopy and scanning electron microscope (SEM), we gain valuable insights into the underlying mechanisms. Additionally, by measuring permeability changes and establishing a correlation between normalized permeability and calcium carbonate content, we shed light on the practical implications of MICP in controlling the permeability of marine soils, particularly in soil surrounding piles. This research contributes significantly to our understanding of MICP behavior in complex marine environments and paves the way for optimizing its application in various engineering scenarios.

# Materials and methods

## Bacteria and cementation solution

In this study, *S. pasteurii* was employed for the experiments. The method (Wang et al., 2019b) was adopted for bacterial culture. The bacterial suspension (775 ul) obtained from the activation of freeze-dried *S. pasteurii* powder was mixed with 80% glycerol (225 ul autoclaved) to prepare a glycerol stock, which was then stored at -80°C. After thawing, bacteria from the glycerol stock were grown on (ATCC) 1376 $NH_4$-YE agar medium at 30°C for 48 hours. Single colonies were picked from the agar medium and transferred to (ATCC) 1376 $NH_4$-YE liquid medium for growth at 30°C and 200 rpm for 24 hours to obtain the initial bacterial solution with $OD_{600}$=1.0. The initial urease activity was 34.9 mM/h, and the initial pH was 8.56. To simulate the process of microbial mineralization in marine environment, artificial seawater (with a salinity of about 3.5%) was prepared using ASTM D1141-98 (2013) (ASTM, 2013), and its composition is shown in Table 1. In this study, deionized water and artificial seawater were used to configure the cementation solution. The components of the cementation solution are shown in Table 2.

## Microfluidic chip experiments and MICP treatment

The structure and preparation method of microfluidic chip used in this study followed previous study (Wang et al., 2019a). The PDMS base and curing agent were mixed and poured onto a silicon plate with a specific pore structure (Wang et al. 2019a), which was then placed in an oven at 65°C after curing for 5 h. After curing, the PDMS was peeled from the silicon plate and punched. Finally, the PDMS and glass plate were sealed into a complete microfluidic chip by exposing them to air plasma.

The experimental setup (shown in Figure 1) included an injection pump, a microfluidic chip, a microscope, a temperature-controlled water bath, and a glove box. The microfluidic injection pump was a Lange LSP01-1A, with a flow control range of 0.001-43.349 ml/min. The microscope was Carl Zeiss Axio Observer 7 from Germany. To control the temperature, the microfluidic chips were placed in a circulating temperature-controlled water bath from Polyscience, which had a temperature control range of -30 ~ 170°C and temperature control accuracy of 0.04°C. To create a low-oxygen environment, the microfluidic chips were placed in a glove box model Vigor FG1800/750TS-ZD. The differential pressure sensor used to measure the pressure difference between the inlet and outlet of the microfluidic chip had a maximum range of 3.5 MPa and an accuracy of 0.25%.

The study consisted of four experiments, summarized in Table 3. Experiments 1 and 2 investigated the effect of elevated salinity (0-35 psu) on MICP by injecting a cementation solution configured with artificial seawater (35 psu) and deionized water (0 psu), respectively. Experiments 2 and 3 investigated the effect of reduced oxygen levels (200000 ppm-2 ppm) by placing microfluidic chips in an atmospheric environment (200000 ppm) and a glove box (2 ppm), respectively. Experiments 3 and 4 examined the effect of lowered temperature (20-10°C) on MICP by placing microfluidic chips at room temperature (20°C) and in a temperature-controlled water bath (10°C), respectively. Finally, Experiments 1 and 4 assessed the combined effects of elevated salinity, diminished oxygen, and lowered temperature on MICP. For each experiment, microfluidic chips were prepared and filled with deionized water. 11.56 μL (1.25PV) bacterial solution with an $OD_{600}$ of 1.0 was injected into the microfluidic chip at a flow rate of 0.863 μL/min (5.6 PV/h) at 20°C. After 24 hours of settling under the corresponding experimental conditions, the cementation solution was injected six times at a flow rate of 0.863 μL/min, with an interval of 24 hours, and stored in the respective experimental environment. The samples were photographed using an optical microscope and processed for statistical analysis after MICP. Deionized water was injected into the mineralized microfluidic chip at a flow rate of 0.5 ml/min and the difference between the inlet and outlet pressures was recorded after the differential pressure sensor reading had stabilized.

## Statistical parameter of calcium carbonate

In this paper, we utilized the Carl Zeiss Axio Observer 7 from Germany, which is equipped with the capability to automatically adjust the image scale based on objective magnification and lens distance. As a result, measurements of the same calcium carbonate crystal region remain consistent across different magnifications and lens distances. We observed the samples using a 10× objective lens. The equivalent diameter of calcium carbonate crystals was calculated by measuring the area of individual crystals. Additionally, we determined the number of calcium carbonate crystals in a specific region by counting them and measuring the pore volume. To calculate the crystal size, an equivalent diameter formula was introduced:

$$D = \sqrt{\frac{4A}{\pi}} \tag{3}$$

where, $A$ represents the area of the calcium carbonate, and $D$ represents the equivalent diameter.

To calculate the chemical transformation efficiency (CTE) of calcium carbonate, the following formula was used (Wang et al., 2022; Wang et al., 2023c):

$$\frac{\mathrm{V_{c100\%}}}{\mathrm{V_v}} = \frac{0.5\times\mathrm{IN}\times100}{2.71\times1000}\times100\% \tag{4}$$

$$\mathrm{CTE} = \frac{\frac{\mathrm{V_c}}{\mathrm{V_v}}}{\frac{\mathrm{V_{c100\%}}}{\mathrm{V_v}}}\times100\% \tag{5}$$

where, $V_c$ represents the total volume of calcium carbonate generated by MICP, $V_{c100\%}$ represents the total volume of calcium carbonate that should be generated assuming complete reaction of $Ca^{2+}$ in the injected cementation solution, $V_v$ represents the pore volume, *IN* represents the injection number of cementation solution, and *CTE* represents the chemical transformation efficiency.

### Raman spectroscopy

High-resolution micro-Raman spectroscopy (532 nm excitation light, WITec Alpha300R) was utilized to detect calcium carbonate minerals in the microfluidic chip. The calcium carbonate in different shapes was selected for observation using a 50× objective, and Raman scattered spectra were collected by the spectrometer. The Raman shift, defined as the difference in wavenumbers between the incident and emitted light, was used to analyze the material composition. In this paper, the range of 0-1400 $cm^{-1}$ was selected to analyze $CaCO_3$ mineral types.

### Scanning electron microscope (SEM)

The microstructure of $CaCO_3$ in the microfluidic chip after MICP was analyzed using a scanning electron microscope (SEM). In this paper, the polydimethylsiloxane (PDMS) of the chip with $CaCO_3$ precipitation was carefully removed and then dried in a 50°C oven for 48 hours. The dried sample was cut to the appropriate size and placed onto a scanning stage for gold coating under vacuum conditions. Finally, the gold-coated sample was observed and photographed using the FEI Nova NanoSem 450.

# Results and discussion

### Bacterial growth, attachment and detachment at different stages of MICP

During various stages of MICP, the abundance of bacteria is influenced by a range of bacterial behaviors and their interactions with soil particles and pore fluid. Following bacterial injection, the bacteria undergo growth and settling due to gravity surpassing floating force (Wang et al., 2019a). The microscope images of bacterial growth within 24 hours after injection are shown in Fig.2a and the quantification results are shown in Fig.2b. The bacterial growth rate in the four conditions, namely, DI water-20°C-aerobic (Condition 1), seawater-20°C-aerobic (Condition 2), seawater-20°C-anaerobic (Condition 3), and seawater-10°C-anaerobic conditions (Condition 4), are 80.7%, 64.2%, 62.0%, and 28.2%, respectively (Fig.2b). Therefore, among the factors examined, lowered temperature exerts the most significant impact on the reduction of bacterial growth as there is a largest reduction in bacterial growth rate from Condition 3 to Condition 4 (Fig.2b). This is followed by elevated salinity, and finally, diminished oxygen. Lowered temperature further exacerbates growth constraints by inhibiting the activity of enzymes involved in *S.pasteurii*'s fundamental metabolic processes (Wang et al., 2023c), affecting biochemical reactions necessary for growth. Additionally, lowered temperature reduces cell membrane fluidity (Murata et al., 1997), hindering nutrient uptake and metabolic waste discharge. High salinity induces osmotic stress, leading to cellular damage, increased membrane permeability, and leakage of cellular components (Takamatsu et al., 2005; Cheng et al., 2014a). To counteract this effect, *S.pasteurii* expends additional energy to maintain osmotic balance and cellular homeostasis, resulting in reduced energy available for growth and reproduction, consequently influencing bacterial growth. Being an aerobic bacterium (Martin et al., 2012), *S.pasteurii*'s energy production efficiency declines under limited oxygen conditions, consequently limiting the rate of bacterial growth and reproduction.

Following the injection of cementation solution, the bacteria undergo detachment from soil particles due to shearing forces between pore fluid and soil particle surfaces (Wang et al., 2019b). The quantification of the remaining bacterial cells that are attached to the microfluidic chips are essential as they are the bacteria that contribute to MICP. The microscope images of bacterial attachment at different stages are shown in Fig.3 and the quantification results are shown in Fig.4. As the injection number of cementation solution increased, the bacterial number in the microfluidic chips of conditions 1 and 2 exhibited a gradual decline, whereas in conditions 3 and 4, there was a sharp decline after the first injection of cementation solution (Fig.4a and 4b). This indicates that diminished oxygen (conditions 3 and 4) is the most significant factor reducing the bacterial attachment (compared conditions 3 and 4 with conditions 1 and 2). The adsorption capacity of *S. pasteurii* is influenced by the interplay of bacterial flagellar movement, extracellular polysaccharides, and biofilm generated through its metabolic processes (Dunne, 2002; Achal et al., 2009). In an aerobic environment, oxygen availability is abundant, supporting the cellular metabolism of *S. pasteurii*. This facilitates the bacterium's flagellar motility and the production of extracellular polysaccharides and biofilm, thereby promoting its adsorption capacity. In contrast, in anaerobic environments, the only source of oxygen is dissolved in the solution. The reduction in oxygen content hinders the cellular metabolism of aerobic bacteria like *S. pasteurii*, subsequently impacting its flagellar motility and its ability to produce extracellular polysaccharides and biofilm. As a result, the adsorption capacity of the bacterium is inhibited under diminished oxygen conditions.

## Morphology and type of calcium carbonate crystals

Microscopic images of the central region of the microfluidic chips after six injections of the cementation solution are shown in Fig.5 for the different conditions. Calcium carbonate is uniformly generated within the microfluidic chip, exhibiting distinct crystal forms under different experimental conditions (Fig.5). The four observation regions in Fig.5 were selected to zoom in to observe the calcium carbonate crystal morphology (shown in Fig.6), and the Raman shift peaks associated with these crystals are shown in Fig.7. Elevated salinity and lowered temperature play significant roles in shaping the morphology and types of calcium carbonate, whereas diminished oxygen has minimal influence on these aspects. In deionized water systems, the predominant crystal shapes observed are rhomboidal (Fig.6 DI-20-Ae, green box), semi-spherical (Fig.6 DI-20-Ae, red box), and combinations of multiple semi-spherical calcium carbonates (Fig.6 DI-20-Ae, orange box). The Raman shift peaks associated with these crystals in deionized water, namely 152, 278, 712, and 1085 $cm^{-1}$ (Fig.7b I, II), indicate their classification as calcite (Yin et al., 2009; Zhang et al., 2010). In contrast to the deionized water system, in all three artificial seawater systems, the primary shapes of calcium carbonate are semi-spherical (Fig.6 Se-20-Ae, Se-20-An, Se-10-An red box) and columnar (Fig.6 Se-20-Ae, Se-20-An, Se-10-An blue box). Additionally, the presence of a Raman shift peak at 1008 $cm^{-1}$ (Fig.7c II, 6d II) indicates the formation of gypsum in the columnar precipitate (Chang et al., 1999; Prieto-Taboada et al., 2014). Comparing the three samples from the artificial seawater systems, under lowered temperature, the Raman shift peaks of semi-spherical calcium carbonate are observed at 1075 and 1089 $cm^{-1}$ (Fig.7e I), suggesting the presence of vaterite (Yin et al., 2009; Cheng et al., 2017).

Previous studies have demonstrated that MICP in seawater can lead to the formation of various minerals, including calcite, vaterite, aragonite, magnesium carbonate, and hemihydrate gypsum (Dong et al., 2021; Peng et al., 2022; Lin et al., 2023; Zhao et al., 2023). Wang et al. (2019b) suggested that the dissolution and recrystallization of vaterite, following Ostwald's law of crystal growth, lead to the formation of larger vaterite or calcite (Wang et al., 2019b). The inclusion of $Mg^{2+}$ has been shown to impede the growth of rhombic calcite (Noyes, 1962), thereby facilitating the transformation of calcium carbonate crystal shape from rhombic to hemispherical. Notably, the diminished oxygen conditions do not directly influence the precipitation process of calcium carbonate crystals; rather, their impact is indirect and occurs through the modulation of carbonate ion concentration in the solution via the bacterial metabolism. As a result, diminished oxygen conditions have minimal effect on the shape and type of calcium carbonate crystals.

In the MICP process with the combined factors of elevated salinity, diminished oxygen, and lowered temperature, salinity primarily influences the shape of calcium carbonate, while low temperature chiefly determines its type. After introducing elevated salinity, diminished oxygen, and lowered temperature simultaneously, it becomes evident that the morphology of the calcium carbonate (Fig.6 Se-10-An) remains consistent with that observed under conditions of elevated salinity alone (Fig.6 DI-20-Ae, Se-20-Ae), resulting in spherical crystals. This suggests that salinity has the strongest control on calcium carbonate morphology when all factors act simultaneously on MICP. Furthermore, the type of calcium carbonate (Fig.7e) aligns with observations under low temperature conditions alone (Fig.7d, 7e), specifically yielding spherical vaterite. This suggests that temperature has the strongest control on calcium carbonate type when all factors act simultaneously on MICP.

## Single calcium carbonate crystal structures

In deionized water, a three-layer structure of calcium carbonate forms, with traces of bacterial activity observed on the surface of each layer (Fig.8 DI-20-Ae, I). The cross-section of calcium carbonate crystals exhibits two distinct regions. (Fig.8 DI-20-Ae, II). In seawater condition, the formation of a six-layer structure of calcium carbonate (Fig.8 Se-20-Ae, I) and solid calcium carbonate (Fig.8 Se-20-An, I) are observed. Additionally, under the corresponding experimental conditions, acicular calcite with a three-layer structure was formed. (Fig.8 Se-20-Ae, II; Se-20-An, II). At 10°C, calcium carbonate crystals with cavities in the center were observed (Fig.8 Se-10-An, I), along with the presence of clusters of acicular calcite (Fig.8 Se-10-An, II). Furthermore, crystal morphology analysis revealed that rhomboidal crystals corresponded to calcite, while hemispherical crystals were identified as vaterite, consistent with previous studies (Wang et al., 2019a; Wang et al., 2019b; Wang et al., 2023c). This classification was further supported by Raman spectroscopy conducted by Xiao et al. (2021) in deionized water, where rhombohedral crystals were confirmed as calcite and hemispherical crystals were identified as vaterite (Xiao et al., 2021).

According to the nucleation theory (Zhang et al., 2002; Tong et al., 2004; Yu et al., 2004), the formation of a new nucleus of calcium carbonate requires overcoming activation energy. The formula for activation energy is shown inequation (6).

$$\Delta G_N = \frac{16\pi\left(\Delta G_1\right)^3}{3\left(kT\ln S\right)^2} \tag{6}$$

$$S = \frac{C\left(Ca^{2+}\right)\times C\left(CO_3^{\,2-}\right)}{K_{sp}} \tag{7}$$

where, $\Delta G_1$ is the surface energy that was needed to form the new interface and maintain the crystal growth, $k$ is the Boltzmann constant, $T$ is the temperature, $S$ is the supersaturation ratio of area, $C\left(Ca^{2+}\right)$ is the calcium ion concentration, $C\left(CO_3^{\,2-}\right)$ is the carbonate ion concentration, and $K_{sp}$ is the solubility product constant of calcium carbonate.

Following the injection of the cementation solution containing $Ca^{2+}$ and urea, which are hydrolyzed into $CO_3^{2-}$, an elevated supersaturation ratio is achieved. This decrease in activation energy required for nucleation triggers the formation of new nuclei on the layer surface (Equation 6). As the reaction progresses, the concentration of $Ca^{2+}$ and $CO_3^{2-}$ gradually decreases, resulting in a decline in the supersaturation ratio of the solution. Consequently, the activation energy required for nucleation increases, prompting the continued growth of calcium carbonate along the existing crystal surfaces.

## Diameter and number of calcium carbonate crystals

To observe the crystal growth throughout the treatment process under the different conditions, microscope images of one pore in the microfluidic chip were taken 24 h after each of the 1-6 injections of cementation solution, as shown in Fig.9. Fig.9 illustrates that the diameter and number of calcium carbonate crystals gradually increased with the number of cementation solution injections, with varying growth rates observed under different experimental conditions.

The quantification of crystals diameter and number is shown in Figures 10-13. Under all conditions, both the diameter and number of calcium carbonate crystals increased following a power function with the increasing number of the cementation solution injections, eventually reaching a stable state (Fig.10a and b). The growth rate of crystal number and total volume initially increased and then decreased as the number of cementation solution injections increased (Fig.11a and b). The maximum growth rate of crystal number was observed after the second to the third injections, indicating the primary nucleation process. The growth rate of $V_c/V_v$ was the lowest under lowered temperature (Fig.11b). Overall, with more injections of the cementation solution, the particle size distribution curve of calcium carbonate shifted to the right, and the median particle size of each group continued to increase (Fig.12a and b). The coefficient of uniformity ($C_u$) of calcium carbonate gradually decreased with more injections, indicating a more uniform particle size distribution for the generated crystals (Fig.13). Lowered temperature plays the most significant role in reducing in crystal diameter (Fig.10a), while diminished oxygen significantly reduces both crystal number and crystal size (Fig.10b).

Diminished oxygen is the primary factor for reducing the crystal number (Fig.10b), consistent with the finding that diminished oxygen reduces the bacterial attachment, thus affecting remaining bacterial number in the microfluidic chip (Fig.4a). (Wang et al., 2021a) found that the bacterial number is a key factor controlling crystal number, and this study supports Wang et al. (2021), illustrating that factors affecting bacterial number also affect crystal number. Lowered temperature hinders bacterial growth, thereby affecting bacterial number. Due to the low number of bacterial cells, the number of crystals is low. Additionally, although lowered temperature does not hinder bacterial specific ureolytic activity (Wang et al., 2023c), it hinders the growth of $CaCO_3$ crystals (Wang et al., 2023c). Therefore, temperature hinders crystal number by affecting bacterial growth, consequently reducing bacterial number, and hinders crystal size by directly affecting the growth of $CaCO_3$.

## Chemical transformation efficiency of calcium carbonate

The overall effects of elevated salinity, diminished oxygen, and lowered temperature on crystal number and crystal size can be represented by chemical transformation efficiency (Fig.14a) and volume proportion of calcium carbonate to pore volume (Fig.14b), indicating the overall $CaCO_3$ content produced. Lowered temperature (10°C) is the most significant factor affecting the overall $CaCO_3$ content. This is consistent with previous sections, which show that lowered temperature affect both crystal number and crystal size. The chemical transformation efficiency at lowered temperature (10°C) was about 10-20% across all injections of cementation solution, whereas the chemical transformation efficiency is about 50-70% for the other three cases (Fig.14b). This results in the low volume proportion of calcium carbonate to pore volume at lowered temperature, which is only about 0.015 after six injections of cementation solution. In contract, for the other three conditions, the values are about 0.06-0.08. After six injections of the cementation solution, the chemical transformation efficiency of MICP-treated samples under combined marine conditions (elevated salinity, diminished oxygen, and lowered temperature) is 20.6% of that in DI water at atmospheric oxygen level and 20°C, with lowered temperature being the primary contributor to this reduction (40.1%).

Additionally, total bacterial activity determines the chemical transformation efficiency of calcium carbonate. Elevated salinity and lowered temperature primarily influence the chemical transformation efficiency by impacting the specific bacterial activity, whereas diminished oxygen mainly affects it by altering bacterial quantity. After six injections of the cementation solution, both the elevated salinity group (Fig.4a DI-20-Ae, Se-20-Ae) and the lowered temperature group (Fig.4a Se-20-An, Se-10-An) exhibited similar bacterial quantities. However, their chemical transformation efficiency decreased by 5.6% and 40.1%, respectively. Urease activity has been shown to decrease by 5.4% immediately after mixing bacteria with seawater (Cheng et al., 2014b). Lowering the temperature from 20°C to 10°C results in a reduction of urease activity by approximately 50% (Van Paassen, 2009; Wang et al., 2023c). Interestingly, the extent of specific bacterial activity reduction due to elevated salinity and lowered temperature closely matches the degree of decrease in chemical transformation efficiency attributable to these same factors. Consequently, in cases of similar bacterial quantity, the inhibition of bacterial specific activity by salinity and low temperature, which ultimately leads to a decrease in total bacterial activity, is the main reason for the limitation of calcium carbonate production by salinity and low temperature. In the group with diminished oxygen (Fig.4a Se-20-Ae, Se-20-An), there was a notable decrease in the bacterial quantity, accompanied by a 9.1% reduction in the chemical transformation efficiency. Oxygen content had essentially no effect on urease activity (Whiffin, 2004; Jiang et al., 2016; Lapierre et al., 2024). Therefore, lower bacterial numbers leading to lower total bacterial activity at similar specific activity is the main reason why diminished oxygen limit calcium carbonate production.

## Effect of calcium carbonate on permeability of the porous medium

The normalized permeability demonstrates an exponential decline as the calcium carbonate content increases following two stages: fast decay and slow decay (Fig.15). These findings align with previous studies (Whiffin et al., 2007; Yasuhara et al., 2011; Al Qabany et al., 2013; Martinez et al., 2013; Zamani et al., 2019). According to Pore Network Model (PNM) theory, the pores in microfluidic chip can be categorized into two types based on their sizes: pores (Fig.15b I Pore) and throats (Fig.15b I Throat). The injected bacteria and cementation solution tend to concentrate at the contact points between particles (DeJong et al., 2010), where calcium carbonate subsequently forms at the pore throats. Compared to the pore spaces, the throats are smaller, and the generated calcium carbonate occupies the primary flow channels within the pores, significantly impacting the permeability. The initial generation of calcium carbonate at the throats leads to a swift reduction in permeability (Fig.15b II). As calcium carbonate forms, the cross-sectional area of the throats decreases. Consequently, at the same injection flow rate, the fluid velocity through the throats increases, resulting in lower concentrations of bacteria and cementation solution at the throats. A considerable portion of the bacterial cementation solution is flushed into the pore spaces, leading to calcium carbonate generation within them. As the calcium carbonate content increases, it forms on the inner walls of the pores, increasing the surface roughness and causing an elevation in fluid flow resistance. During this phase, the generation of calcium carbonate within the pores leads to a gradual decline in permeability (Fig.15b III). When the calcium carbonate content continues to rise, any throats or pore spaces within the flow path can become completely obstructed by calcium carbonate, rendering the condition incapable of flowing. Although microfluidic chip technology cannot fully simulate the flow field distribution and spatial distribution of generated calcium carbonate in the 3D structure of the real soil skeleton, pores in real soil can also be categorized into pores and throats according to PNM. Based on the 2D PNM of the microfluidic chip, the mechanism proposed in this paper for how calcium carbonate content affects permeability offers valuable insights and guidance for understanding its impact in three-dimensional real samples. In addition, microfluidic chip technology offers the advantage of real-time visualization of MICP growth under various environmental factors and quantification of bacterial quantity and crystal characteristics (diameter, size, phase change, distribution, growth). In the future, by further improving and optimizing microfluidic chip design and integrating characterization techniques such as Micro-CT, we can more accurately observe the microscale flow and chemical reactions in the real soil skeleton under different environmental factors. This will deepen our understanding of the MICP cementation mechanism and its impact on permeability under various environmental conditions. At the same time, it is possible to domesticate bacteria with high urease activity (e.g. *S. pasteurii*) under extreme conditions to improve the efficiency of microbial mineralization under extreme conditions providing more reliable theoretical support for the application of MICP in marine engineering.

## The "induced" characteristics of MICP and its implication for MICP applications

Microfluidic chip and optical microscopy observation experiments offer invaluable insights into the kinetics of MICP, revealing both the potential and limitations of this process as an induced, rather than controlled, biomineralization method. Through these experiments, it becomes clear that while bacteria can actively initiate the conditions necessary for calcium carbonate formation, such as increasing the pH level of the microenvironment (Zehner et al., 2020), the precise locations and patterns of these mineral deposits remain unpredictable. For instance, it was observed that micrometer-sized precipitated crystals did not grow on negatively charged cell surfaces nor on other tested polystyrene microspheres with different negatively charged surface modifications, indicating that a negatively charged surface was not a sufficient property for nucleating the growth of precipitates in the MICP process (Zhang et al., 2018). In addition, although some bacteria adhere to the microfluidic chip surfaces, many are washed away by the cementation solution (Wang et al., 2019a), and carbonate crystals preferentially precipitate on non-bacterial surfaces, such as the inner walls of the microfluidic chip (Figure 16).

These observations underscore the intrinsic complexity of MICP. The process is indeed inducible—bacterial activity catalyzes the biochemical conditions favorable for mineral deposition. However, the actual deposition sites and the extent of mineral formation are influenced by various factors including microbial distribution, surface properties, fluid dynamics, and environmental conditions within the microenvironments. Microbial density and activity affect MICP morphology because they influence the activation of $CO_3^{2-}$ concentrations, which directly impact nucleation and crystal growth kinetics, consequently affecting MICP crystal number and size (Konstantinou et al., 2021; Wang et al., 2021a; Wang et al., 2022). The high concentration of $CO_3^{2-}$ also contributes to the initial formation of amorphous carbonate precipitates, which can transform into more stable forms of crystals such as vaterite and calcite (Wang et al., 2019b; Wang et al., 2021a). Temperature plays a crucial role in MICP because it affects not only the kinetics of chemical precipitation of $CaCO_3$ but also bacterial growth, density, and attachment (Wang et al., 2023b; Wang et al., 2023c). These factors influence the generation of $CO_3^{2-}$ and the elevation of pH, which in turn affect the precipitation kinetics. In the current study, environmental factors including temperature, oxygen levels, and salinity were found to impact bacterial behavior and the chemical reactions. Because of this complex impact, predicting MICP behavior in real-world environments for different applications (Wang et al., 2023a) when considering all of these factors can be even more challenging. This variability highlights the difficulty of harnessing MICP for targeted applications where precise control over mineralization is necessary (Wang and Konstantinou, 2024).

# Conclusions

This study utilized microfluidic chip technology to conduct microbial mineralization experiments, simulating marine environments characterized by elevated salinity, diminished oxygen, and lowered temperature. By employing measurement techniques such as Raman spectroscopy and SEM, the study examined the influence of elevated salinity, diminished oxygen, and lowered temperature on bacterial growth, adsorption, calcium carbonate chemical transformation efficiency, and chip permeability. Additionally, it provided detailed insights into the effects of these factors on the shape and type of calcium carbonate crystals. These findings hold potential for enhancing the MICP process in marine environments. The specific findings are as follows:

The inhibition of bacterial growth was most pronounced under lowered temperature, followed by elevated salinity and diminished oxygen. Diminished oxygen exerted the important effect in reducing bacterial attachment. Elevated salinity and lowered temperature in seawater environments have a more pronounced effect on the shape and type of calcium carbonate crystals, whereas the impact of diminished oxygen on crystal shape and type is relatively minor.

The presence of $Mg^{2+}$ in seawater inhibits the transformation of vaterite to calcite, and lowered temperature also impedes the rate of vaterite transformation.

Elevated salinity minimally impacts average crystal diameter but reduces crystal number by 20.2%. Diminished oxygen increases average diameter (20.3%) but decrease crystal quantity (50.9%). Lowered temperature decreases average diameter (36.9%) with minimal impact on crystal quantity. After six injections of the cementation solution, the chemical transformation efficiency of MICP-treated samples under combined marine conditions (elevated salinity, diminished oxygen, and lowered temperature) is 20.6% of DI water, atmosphere oxygen level and 20°C, with lowered temperature being the primary contributor to this reduction (40.1%).

The exponential decline of permeability with increasing calcium carbonate content. Based on the location of calcium carbonate generation within the percolation channel, it can be categorized into two types: fast decay and slow decay. These findings hold promising potential for MICP applications for stabilizing soils in marine environments such as soil surrounding piles in the shallow seabed in winter.

The microfluidics investigation in this study underscores the profound impact of environmental conditions on bacterial growth, attachment, crystal growth, and morphology. Although the sizes of the crystals and their chemical transformation efficiency following each of the six injections of cementation solution were quantified, further analyses such as a detailed kinetic study of crystal growth between injections would provide deeper insights into the physicochemical dynamics of MICP. Such a kinetic study would benefit from assessing both bacterial counts using microfluidic chip experiments and bacterial activity through methods such as enzyme assays for urease, qPCR for gene expression, and ATP measurements, all of which would enhance our understanding of the microbes' active roles in the MICP process under varying environmental conditions. Additionally, research into different bacterial strains and chemical interventions that could significantly improve MICP efficiency in challenging environments is warranted. Moreover, long-term stability tests and erosion resistance analyses of the precipitated carbonate are crucial to validate the practical applicability of MICP, particularly in marine and other natural settings. Advancing these research directions will significantly develop the field of MICP to meet the challenges of extreme environments.

Looking forward, enhancing the predictability and control of MICP requires a deeper understanding of microbial interactions with their microenvironments and the development of new techniques to manipulate these interactions. Future research should focus on engineering microbial strains or modifying the microenvironment to direct carbonate deposition more reliably. Additionally, integrating findings from microscale studies like those conducted with microfluidic chips with macroscale applications will be crucial. This integration will help ensure that laboratory insights translate effectively into practical strategies, potentially transforming MICP into a tool for bioremediation, construction, and other fields requiring precise biomineralization processes.

# Data Availability Statement

Data, models, or code that support the findings of this study are available from the corresponding author upon reasonable request.

# Acknowledgments

The authors acknowledge the financial support of National Natural Science Foundation of China (Grant No. 52171262) and Science and Technology Innovation Committee of Shenzhen (Grant No. JCYJ20210324103812033) for conducting this study.

**Table 1** Composition of artificial seawater artificial seawater

| Composition | NaCl | KCl | $NaSO_4$ | $CaCl_2$ | $MgCl_2$ | KBr | $NaHCO_3$ | $H_3BO_4$ |
|---|---|---|---|---|---|---|---|---|
| Content/($g \cdot L^{-1}$) | 24.53 | 0.695 | 4.09 | 1.16 | 5.2 | 0.101 | 0.201 | 0.027 |

**Table 2** Composition of artificial seawater artificial seawater cementation solution (CS)

| Type | $CaCl_2$ | Urea | Nutrient broth | Solvent |
|---|---|---|---|---|
| CS-DI | 0.5 M | 0.75 M | 3 g/L | Deionized water |
| CS-Sea | 0.5 M | 0.75 M | 3 g/L | Seawater |

**Table 3** MICP treatment parameters for the microfluidic chip experiment

| Sample Number | Sample Name | Salinity | Temperature/°C | Oxygen level |
|---|---|---|---|---|
| 1 | DI-20-Ae | DI water (0%) | 20 | Atmospheric level |
| 2 | Se-20-Ae | Seawater (3.5%) | 20 | Atmospheric level |
| 3 | Se-20-An | Seawater (3.5%) | 20 | Low level |
| 4 | Se-10-An | Seawater (3.5%) | 10 | Low level |

## List of Figures

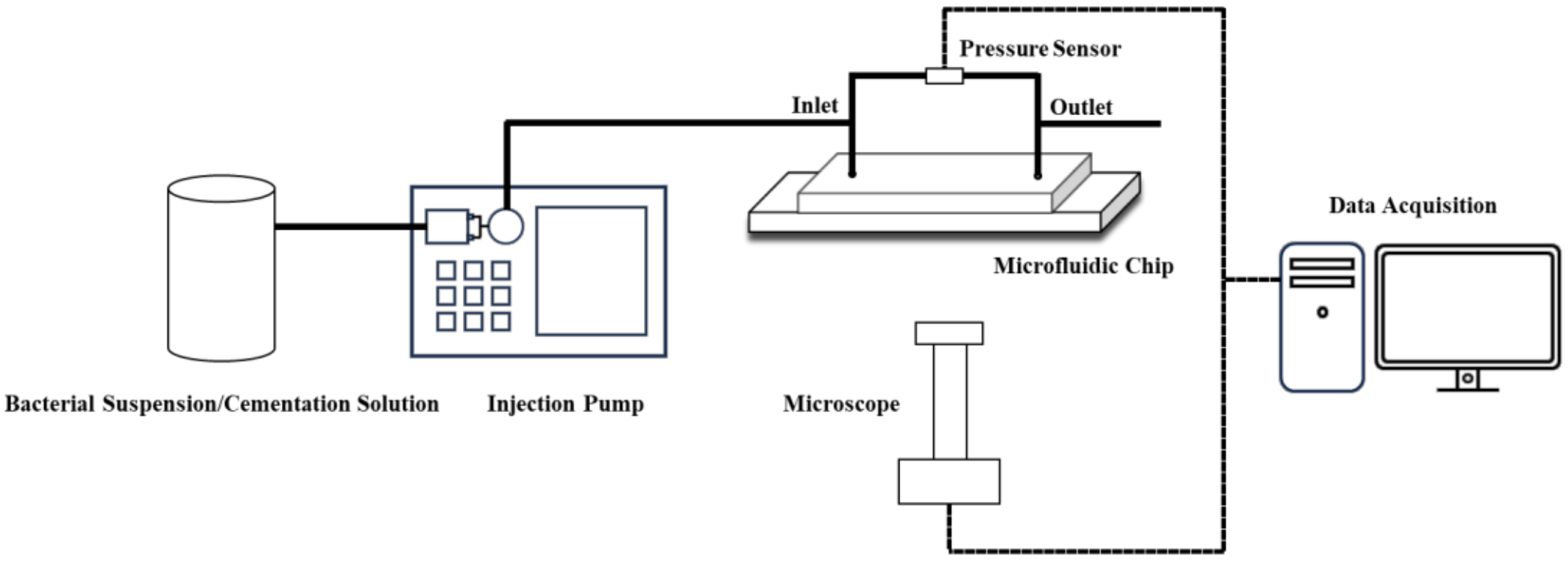

**Fig.1.** Schematic diagram of microfluidic setups

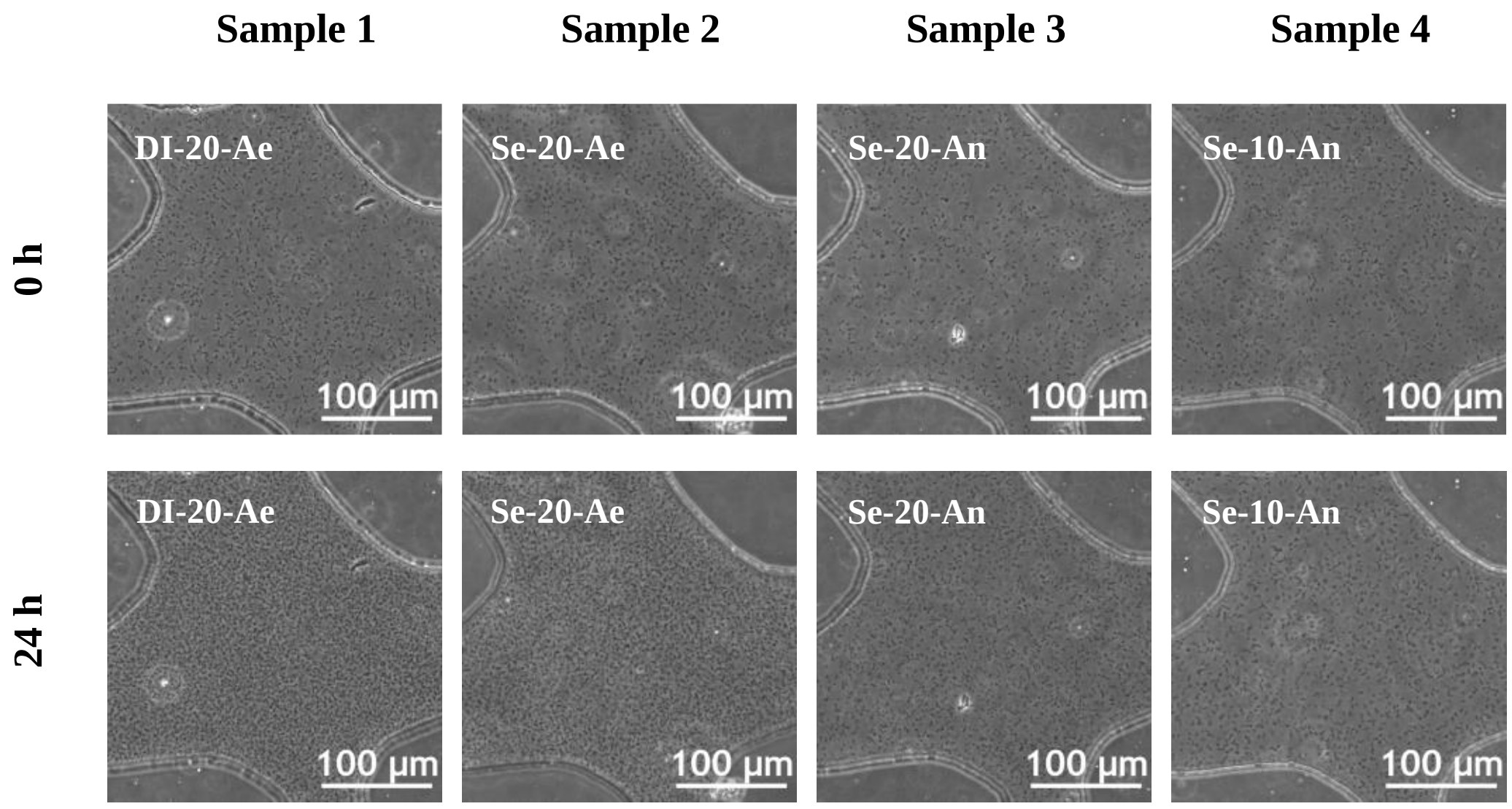

(a)

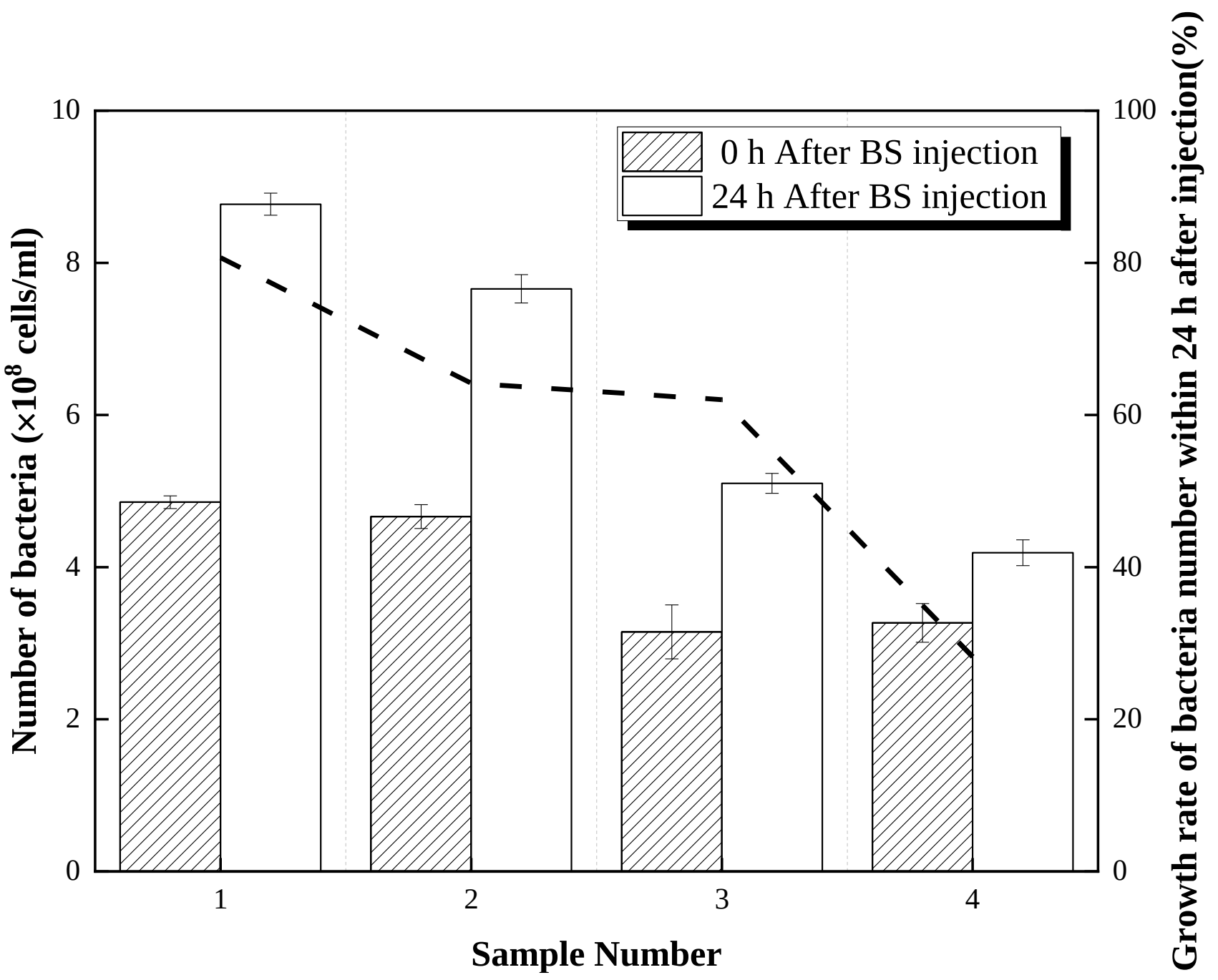

(b)

**Fig.2.** Bacterial quantity at 0 h and 24 h after bacterial injection (a) Microscopic images taken at the middle pore (300 μm × 300 μm) of microfluidic chip; (b) The quantification of bacterial number in the field of observation

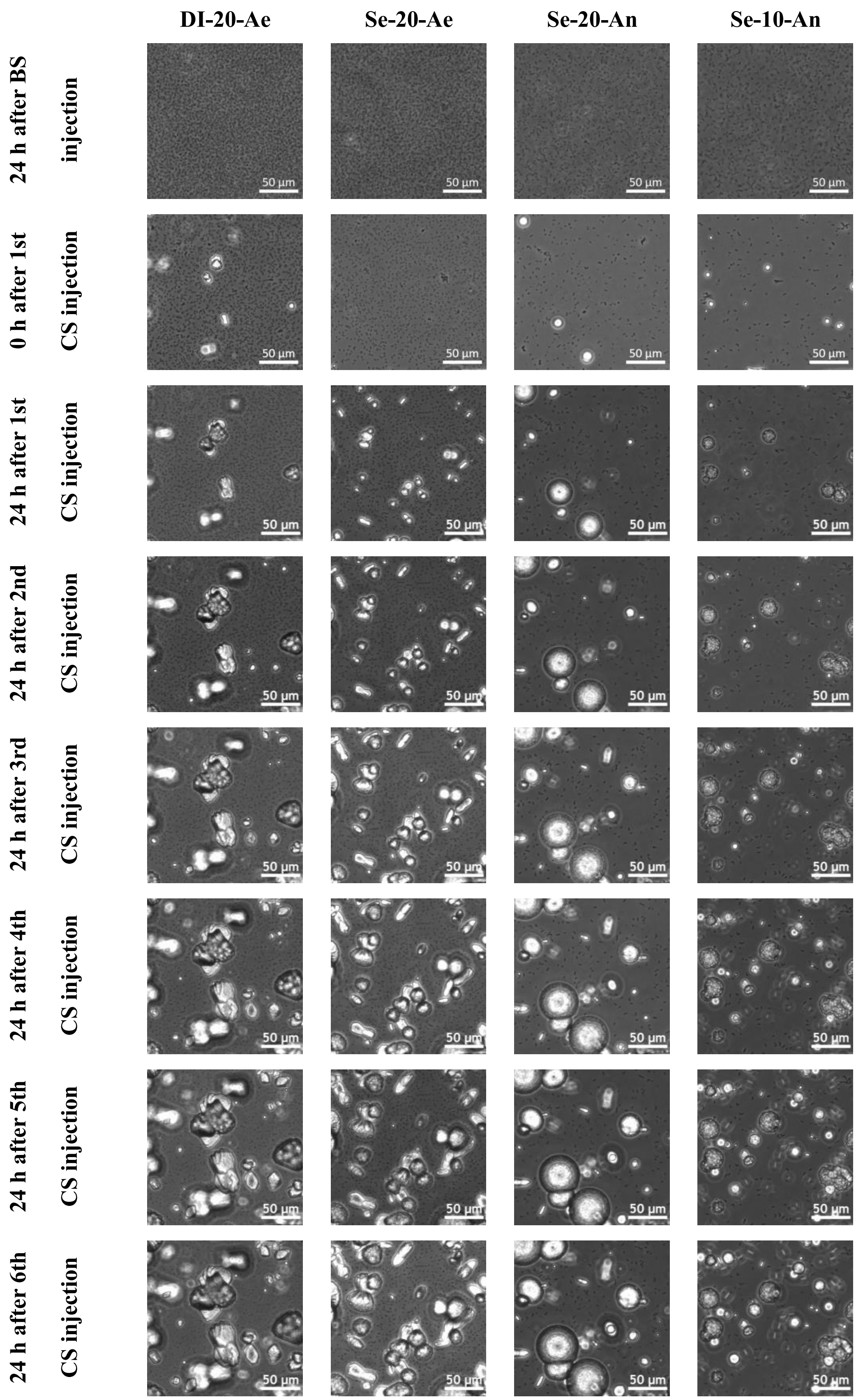

**Fig.3.** Microscope image of microfluidic chip at different stage

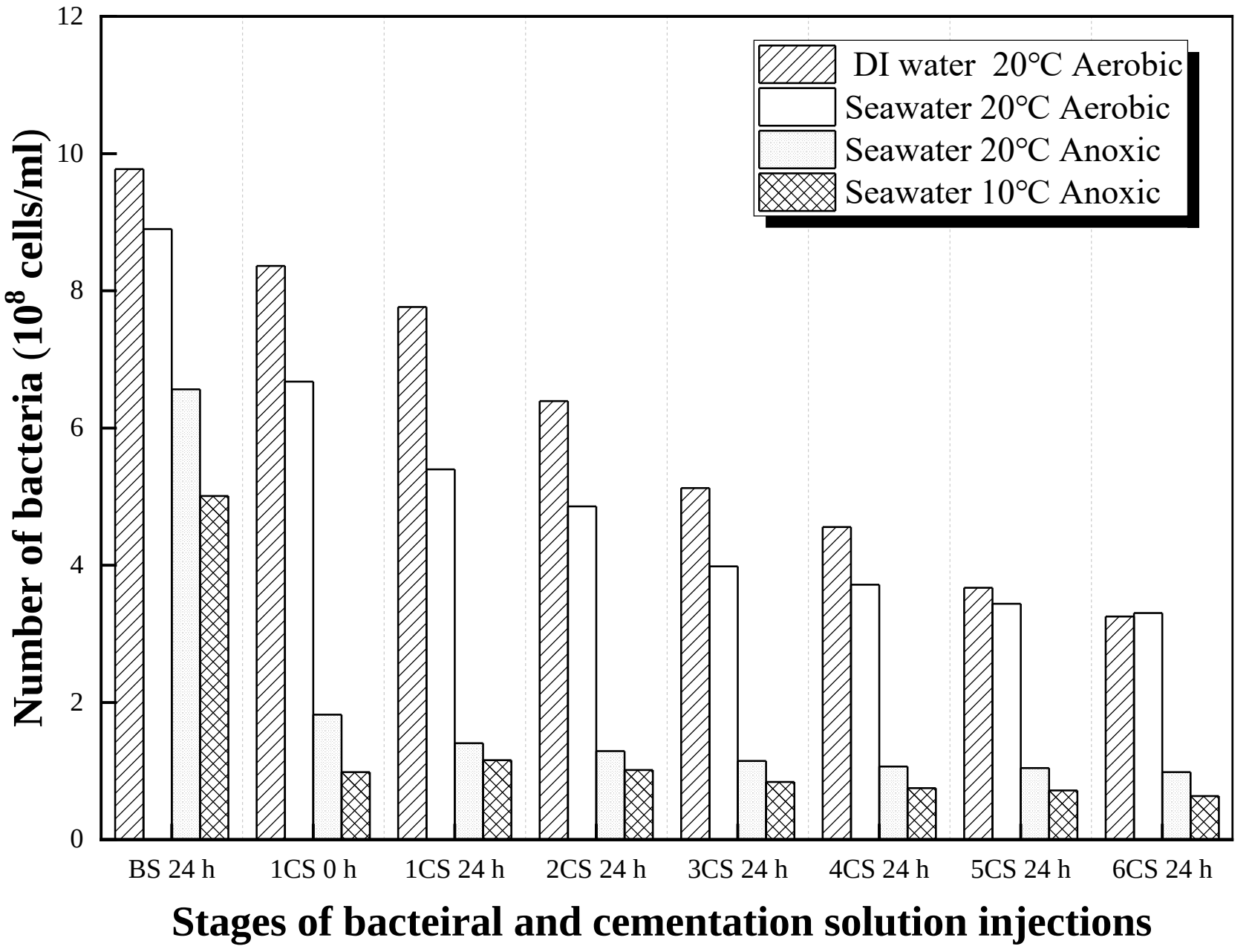

(a)

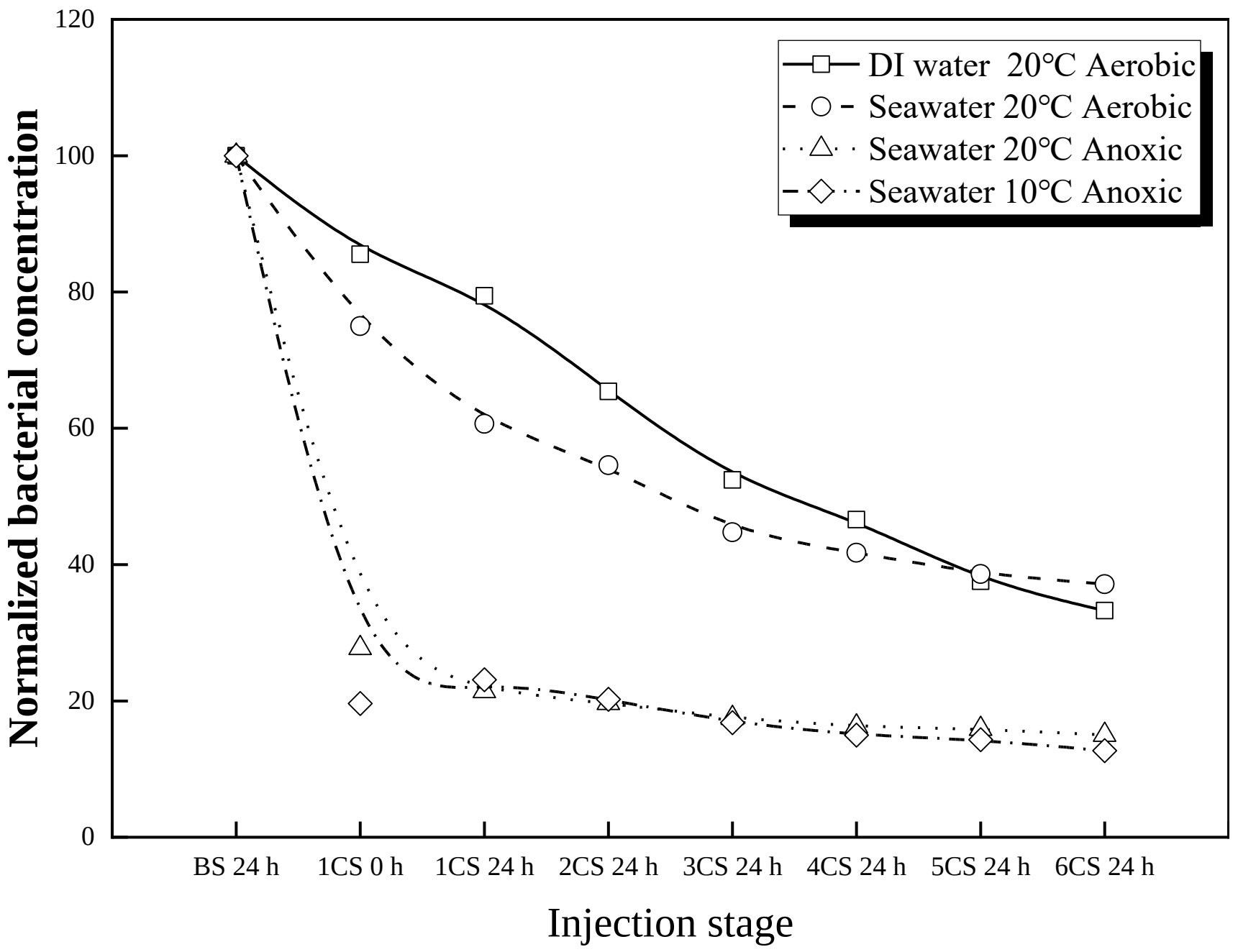

(b)

**Fig.4.** Effect of cementation solution injection number on the number of residual bacteria in microfluidic chip: (a) statistical result of bacteria number; (b) normalized bacterial concentration

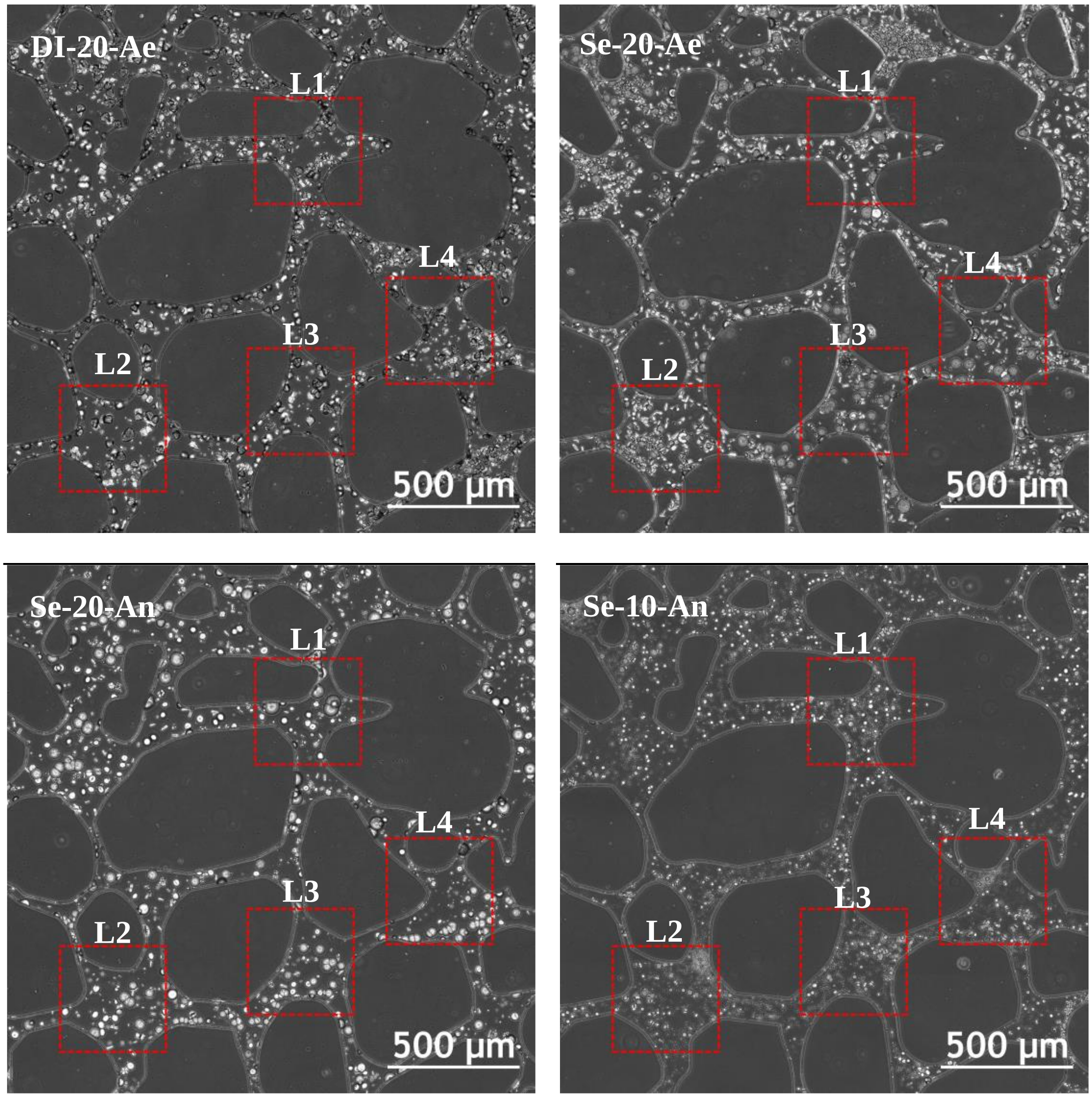

**Fig.5.** Distribution of calcium carbonate crystals after six times of cementation solution injection

(2000 µm×2000 µm)

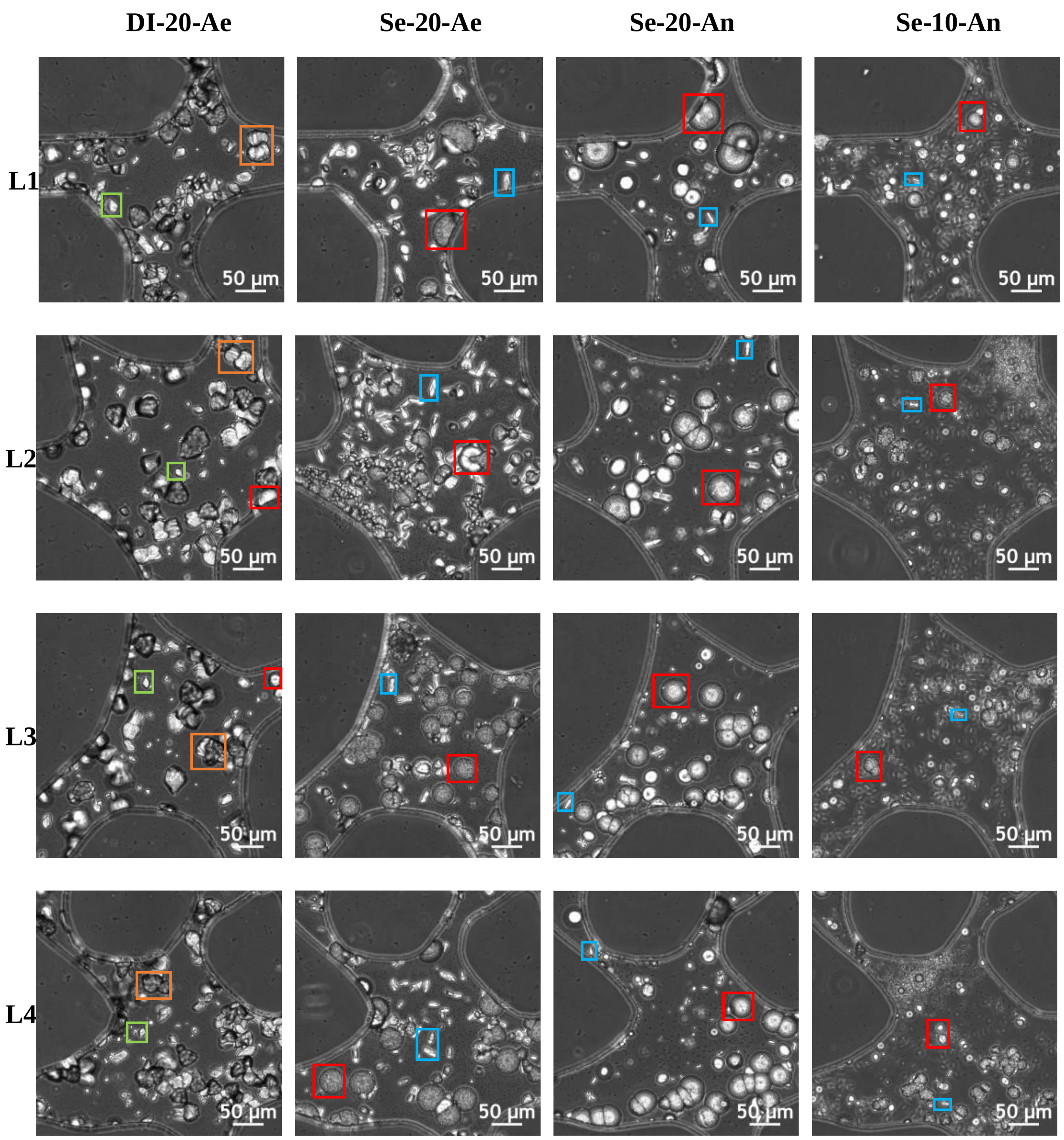

**Fig.6.** Morphology of calcium carbonate crystal after six injections of cementation solution

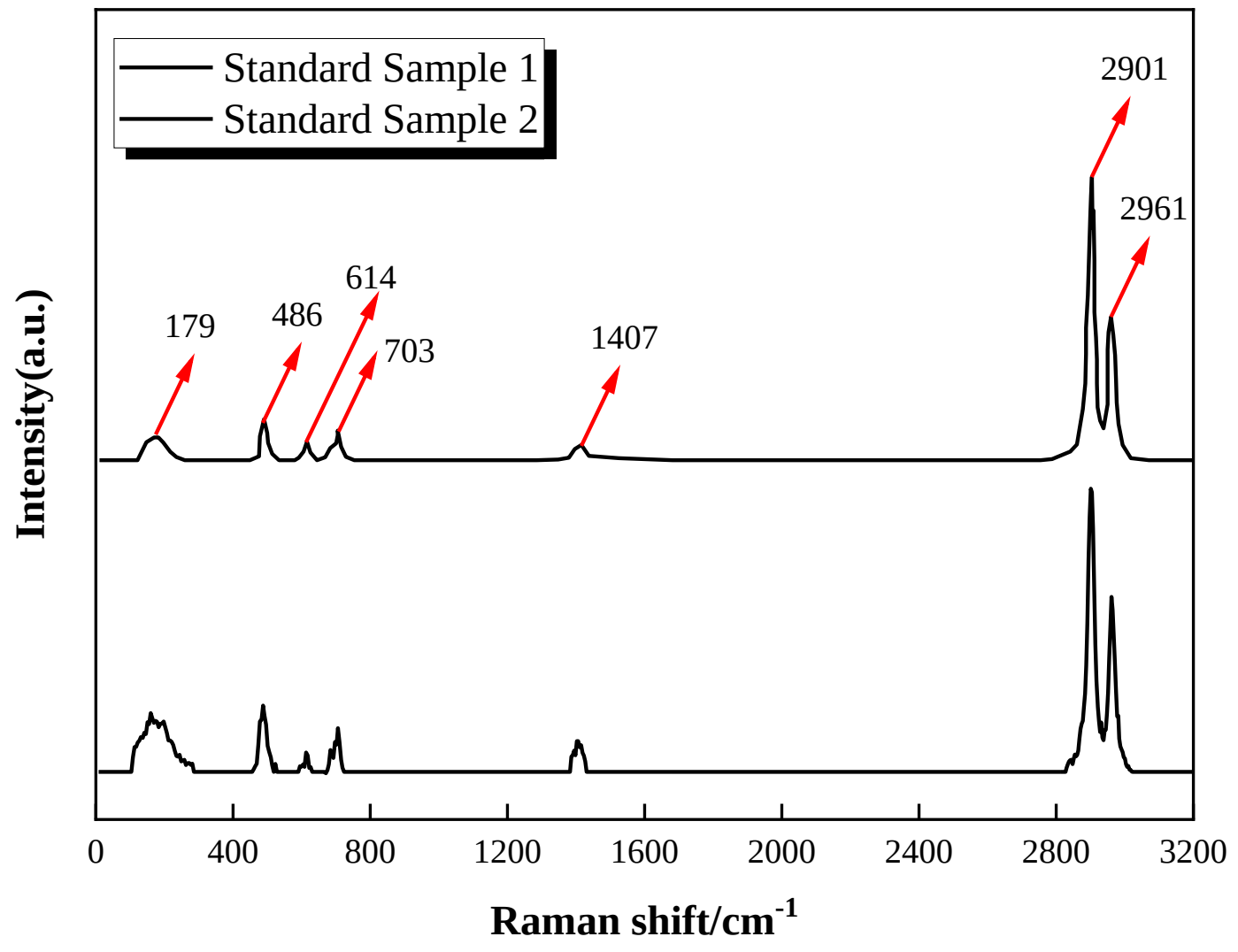

(a)

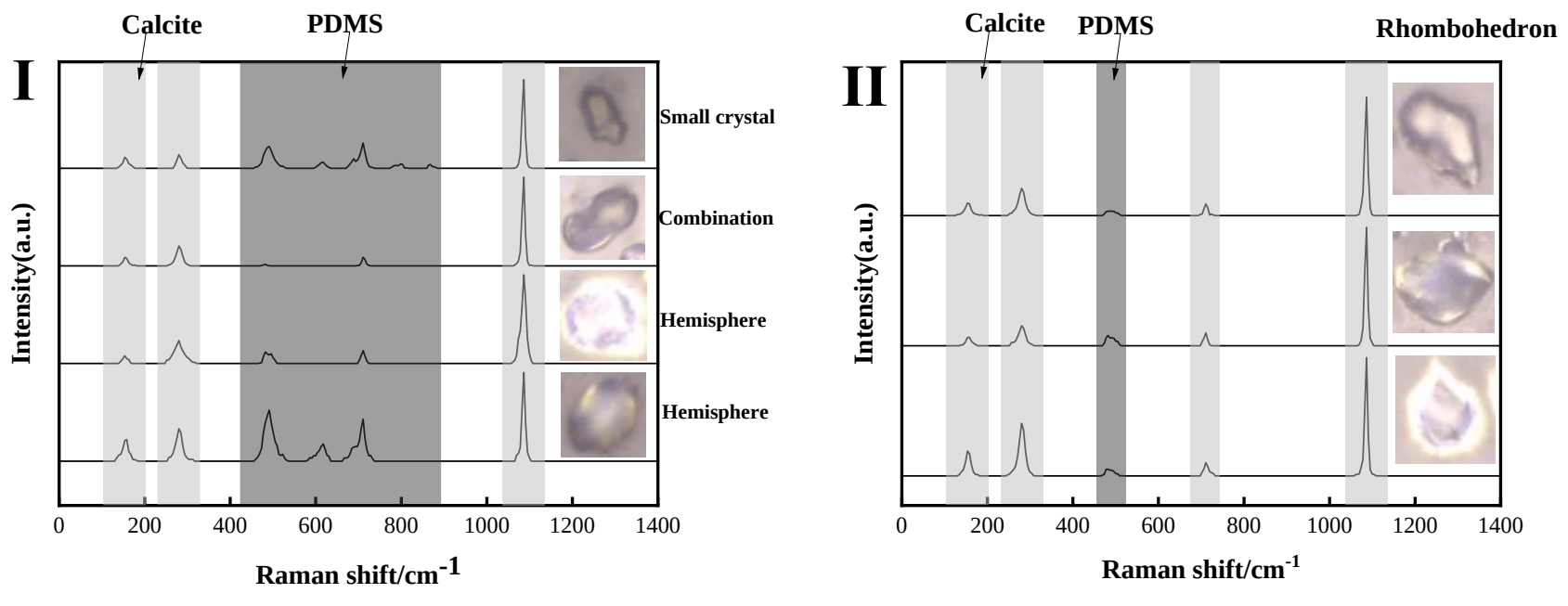

(b)

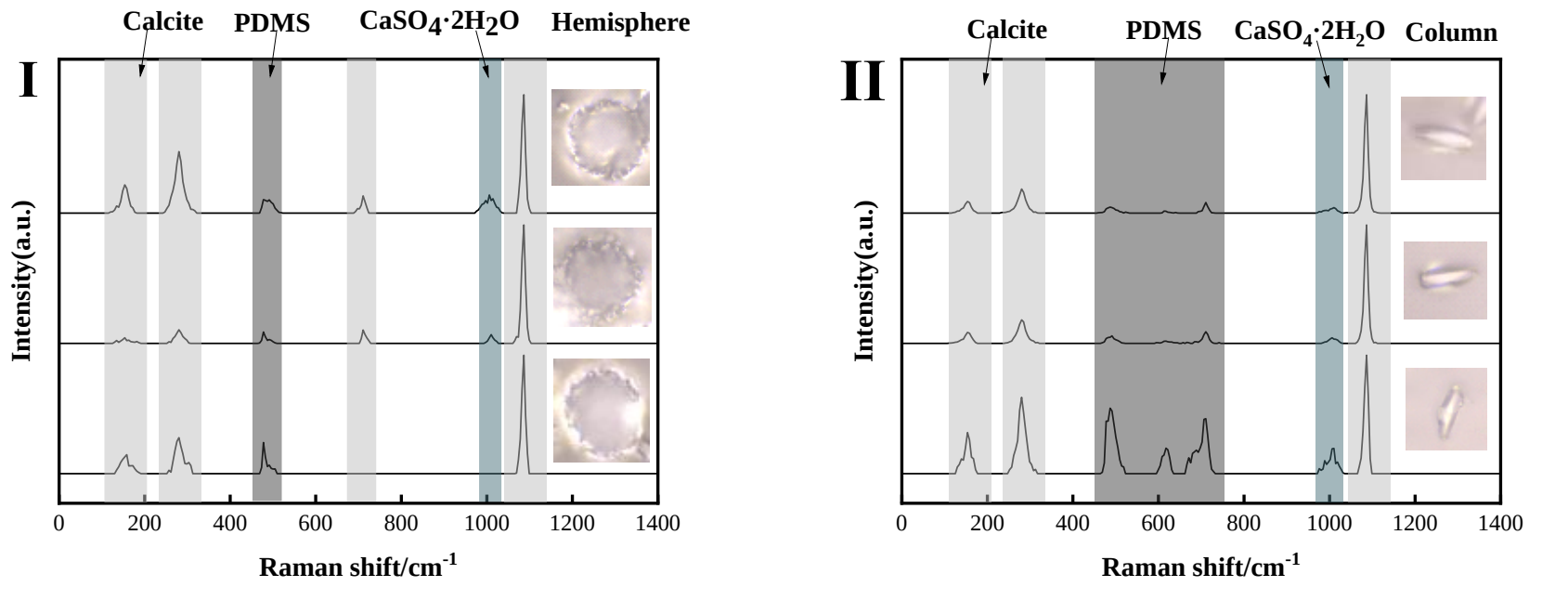

(c)

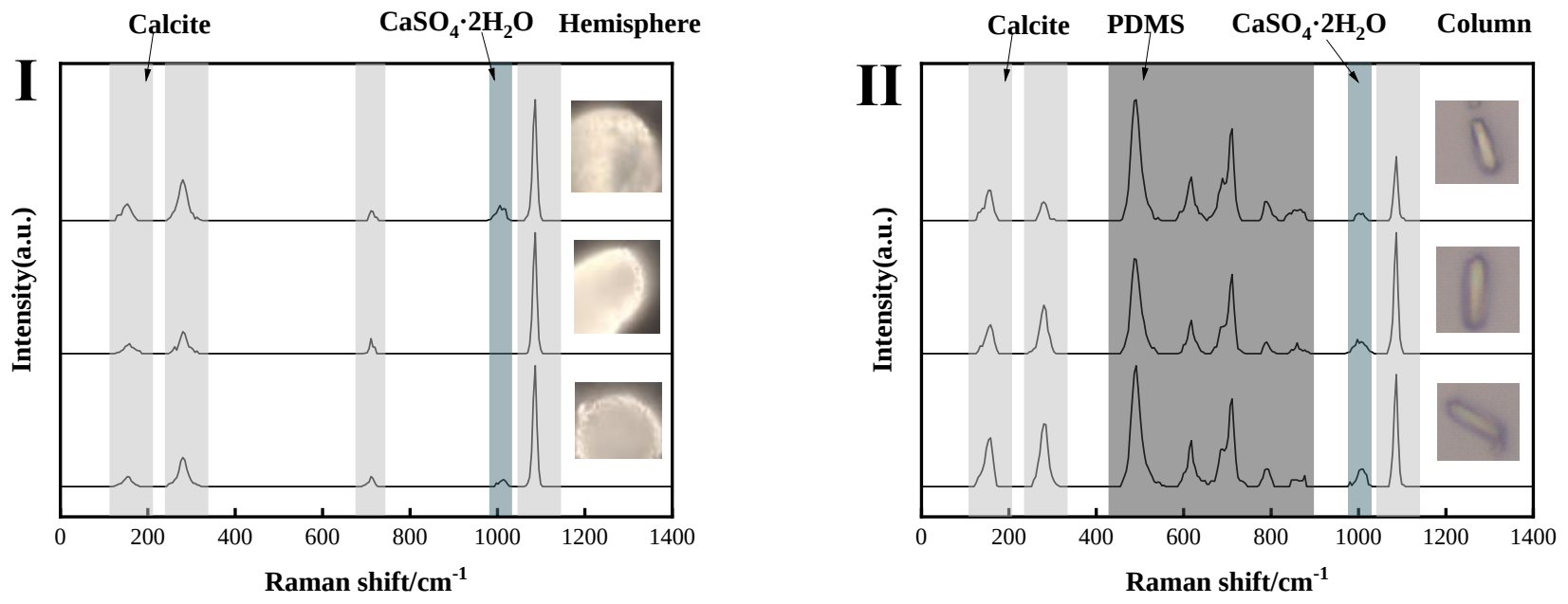

(d)

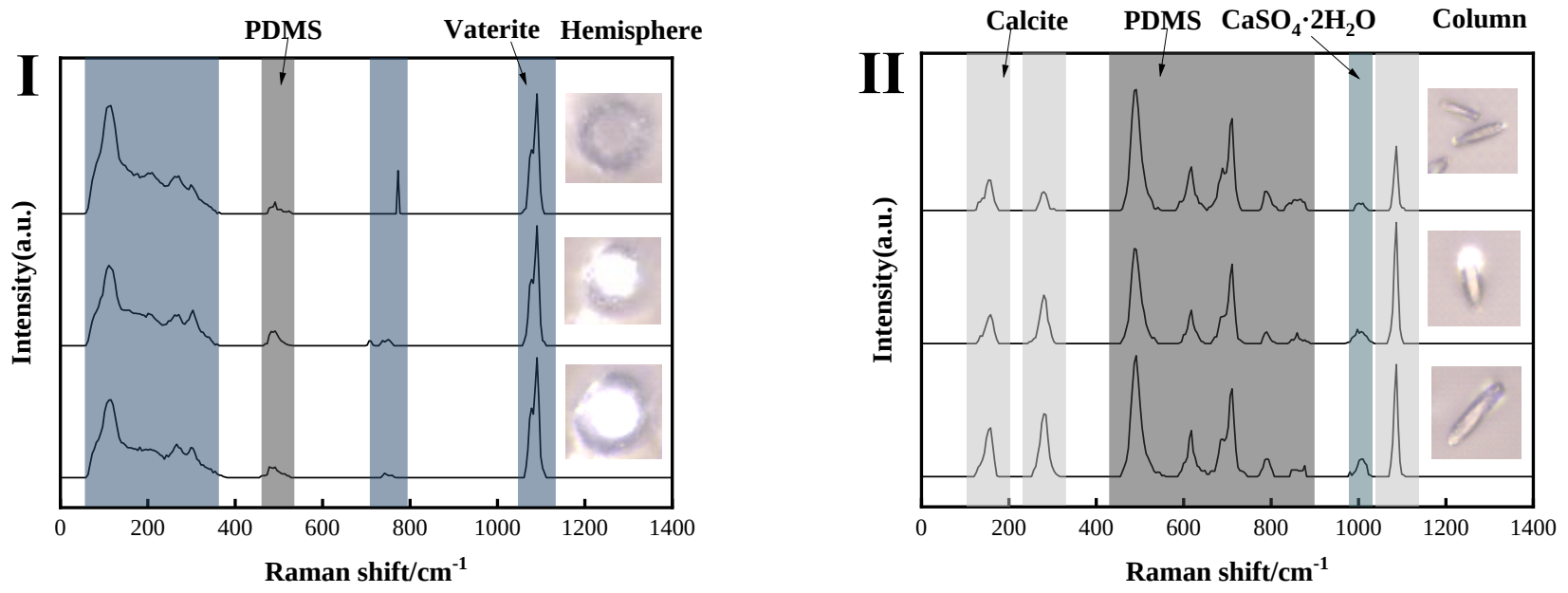

(e)

**Fig.7.** Calcium carbonate Raman spectrum (a) Standard Samples; (b) Sample 1: DI 20°C Aerobic; (c) Sample 2: Seawater 20°C Aerobic; (d) Sample 3: Seawater 20°C Anoxic;(e) Seawater 10°C Anoxic

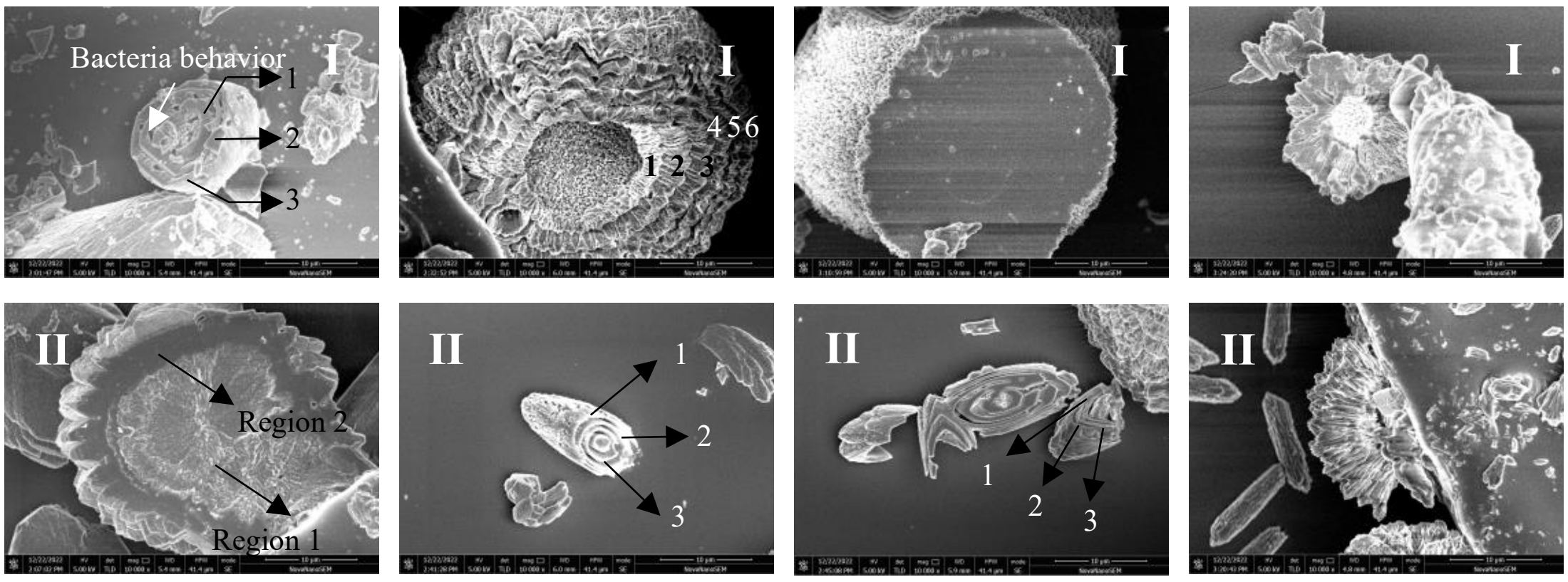

DI-20-Ae    Se-20-Ae    Se-20-An    Se-10-An

**Fig.8.** SEM Images of calcium carbonate crystals presented in microfluidic chips after MICP treatment

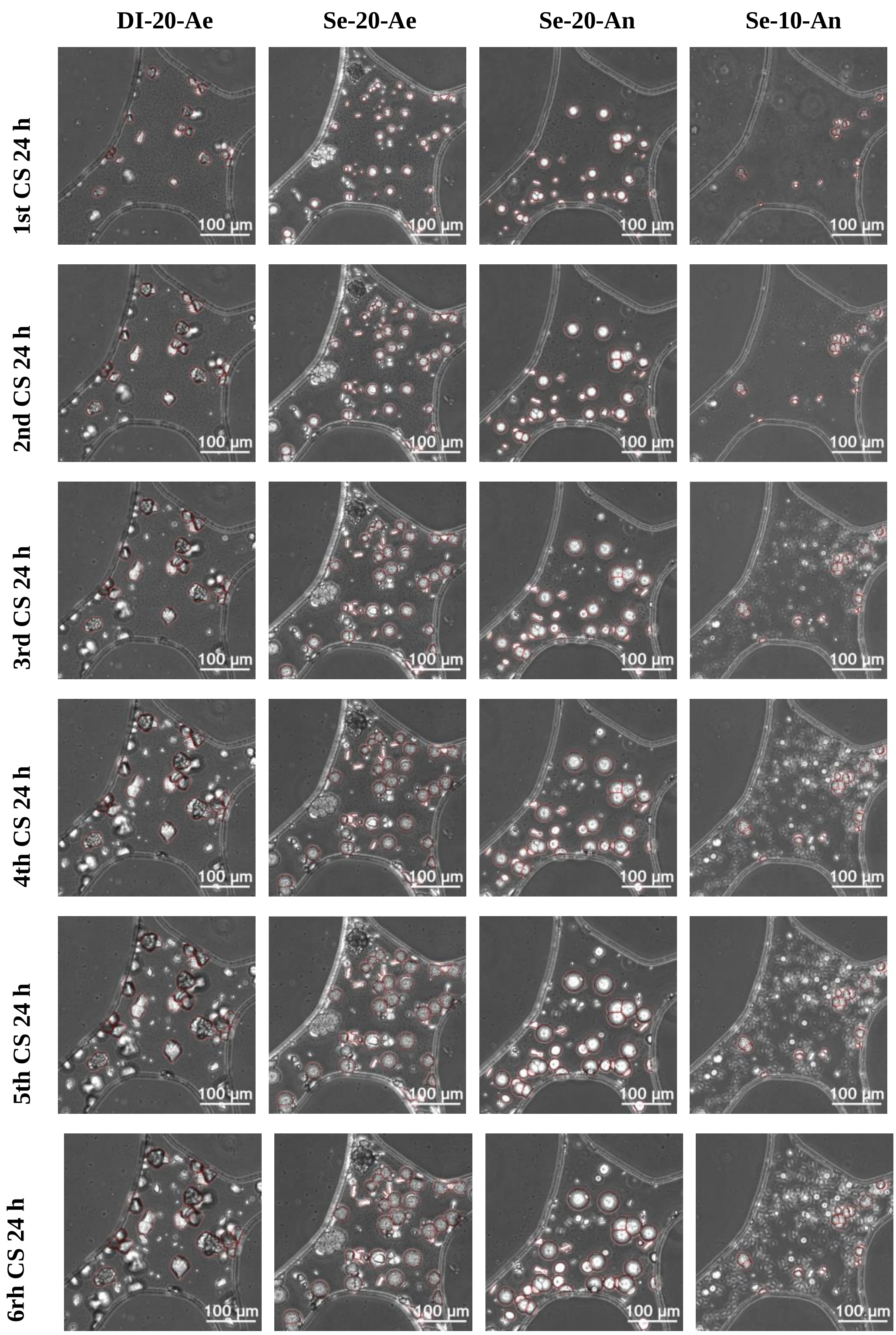

**Fig.9.** Microscopic images of calcium carbonate crystals in a central pore of microfluidic chips taken at 24 hours after 1$^{st}$ to 6$^{th}$ injections of cementation solution

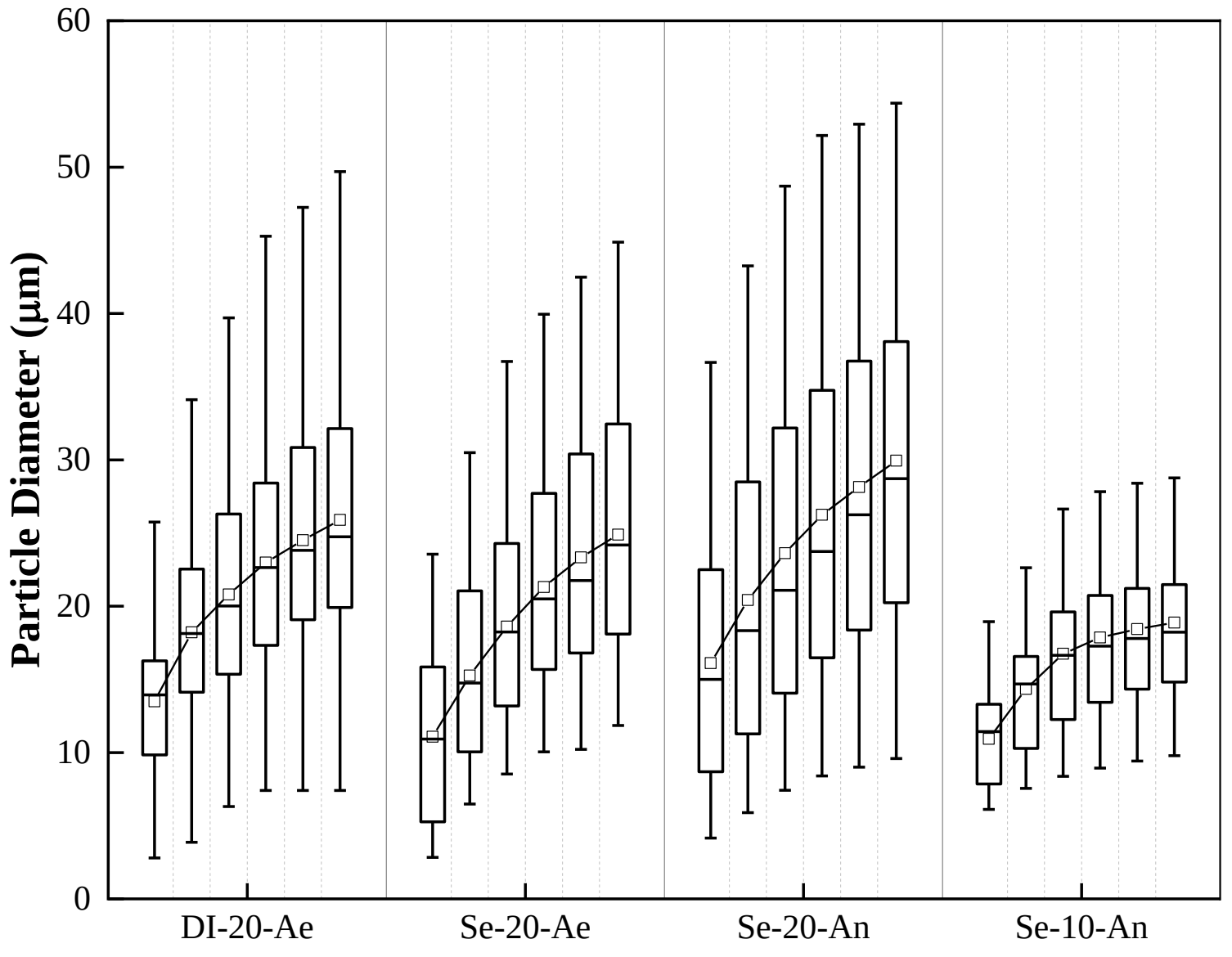

(a)

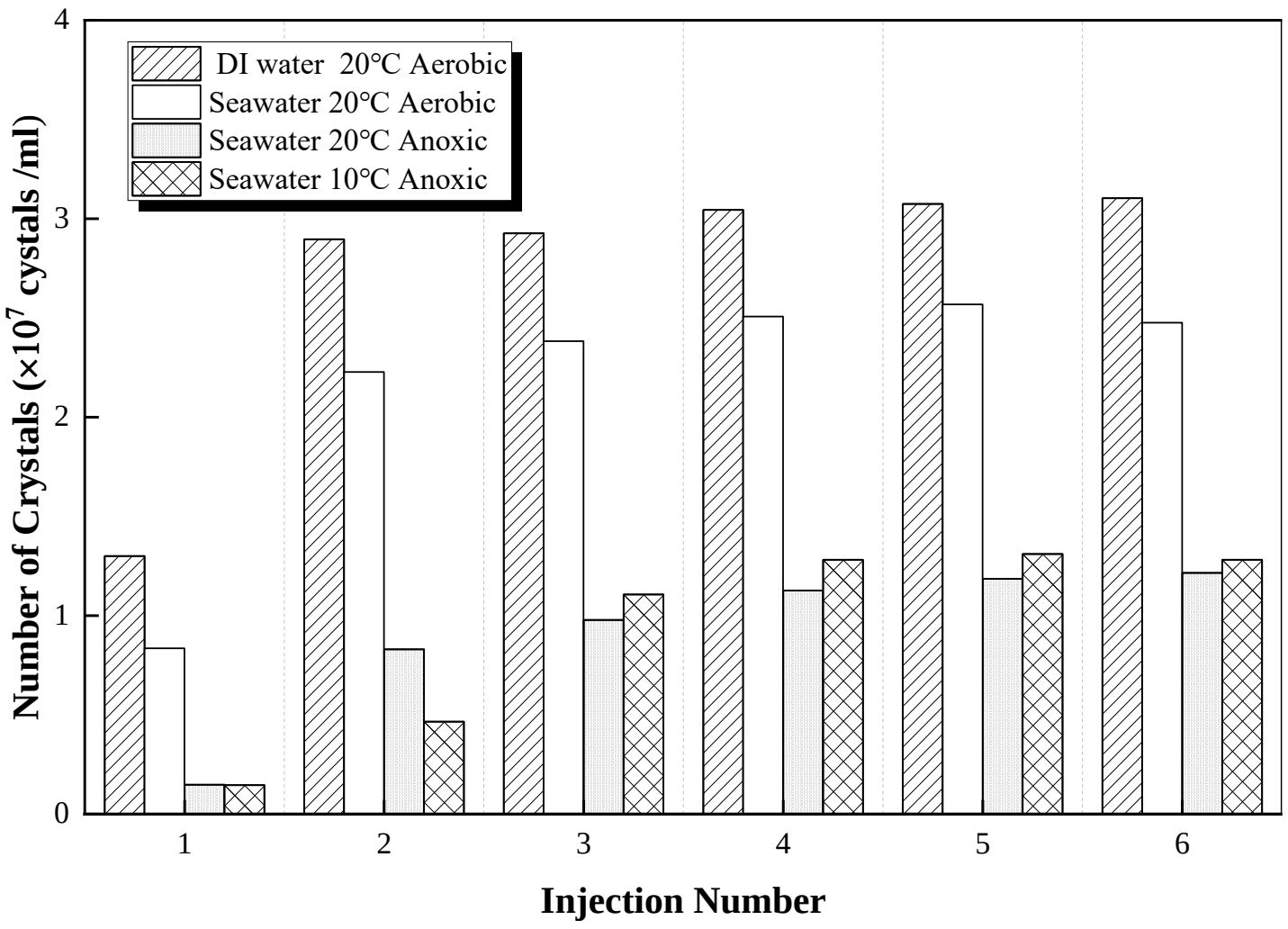

(b)

**Fig.10.** Quantitative analysis of calcium carbonate: (a) diameter of calcium carbonate; (b) number of calcium carbonate

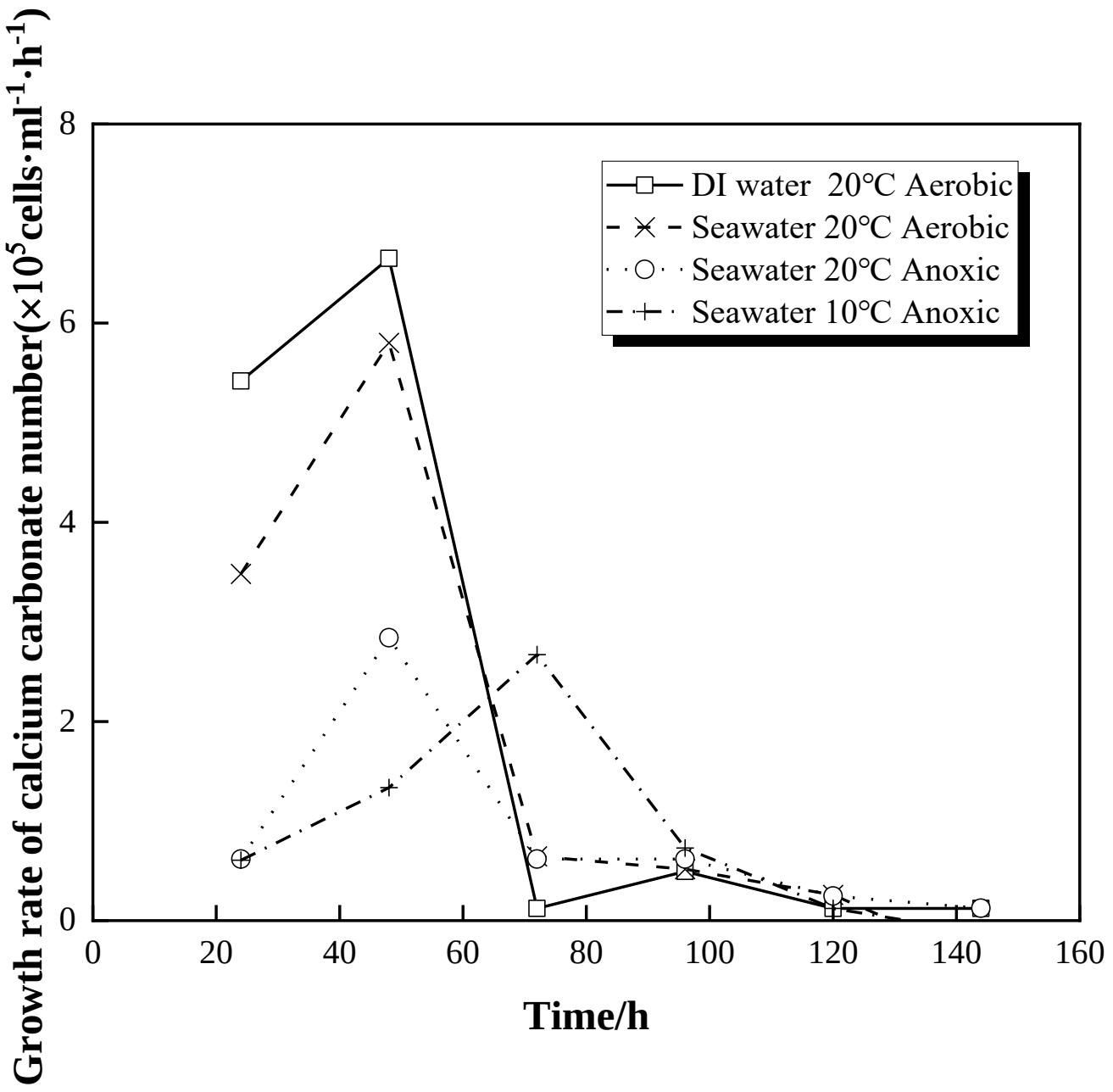

(a)

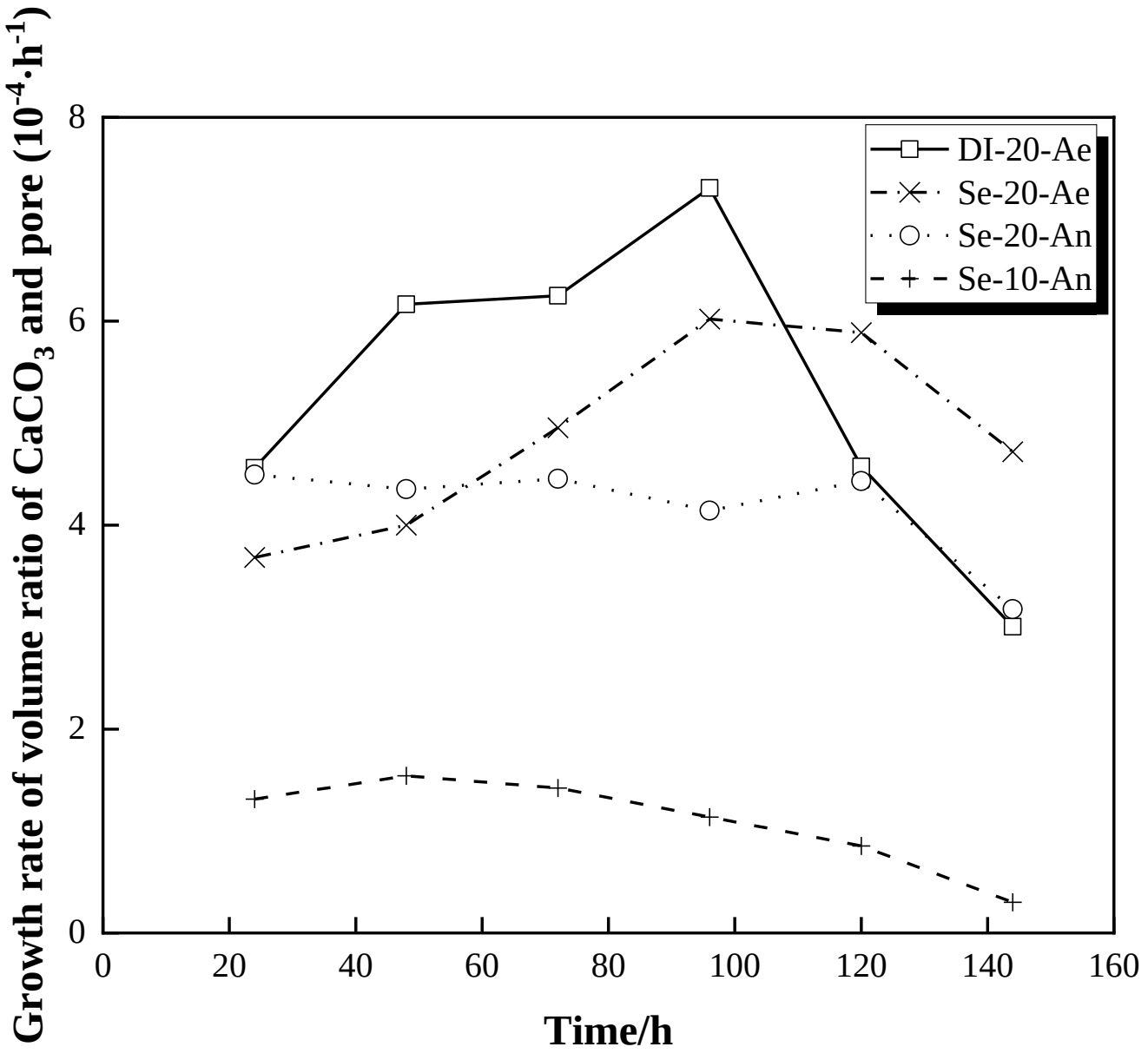

(b)

**Fig.11.** Calcium carbonate growth rate: (a) growth rate of number; (b) growth rate of volume ratio

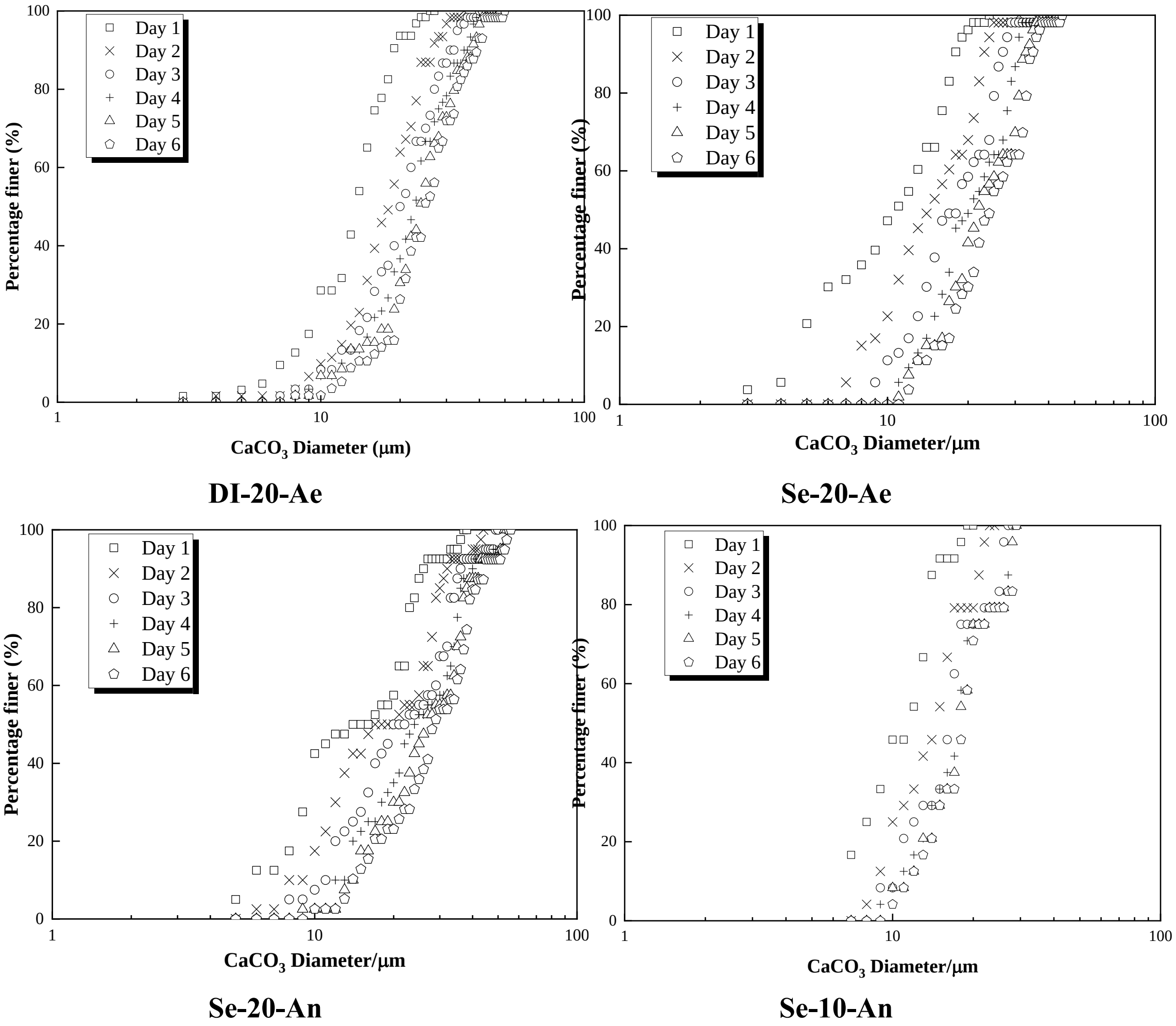

(a)

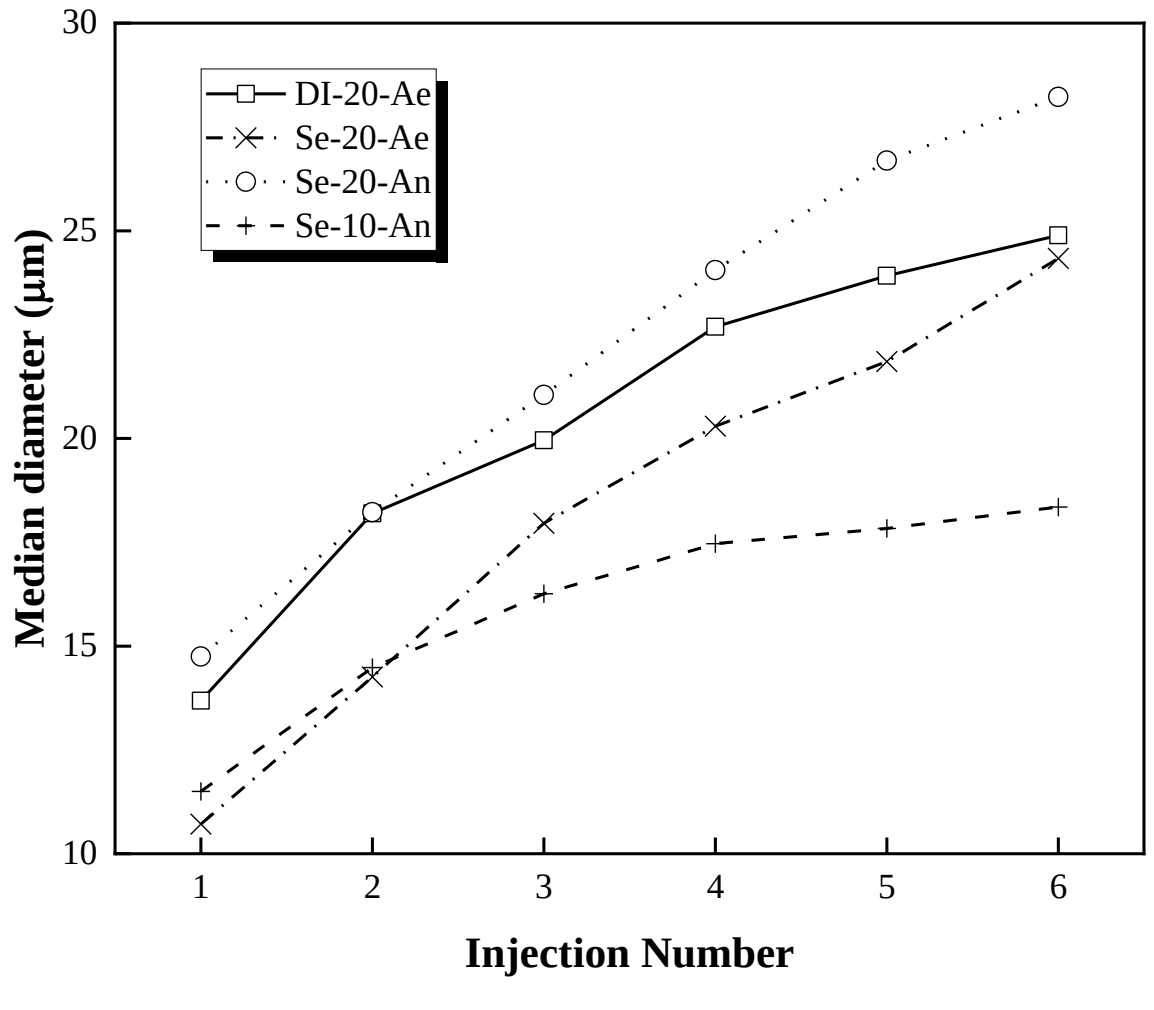

(b)

**Fig.12.** Size distribution of calcium carbonate crystals: (a) grading curve; (b) median diameter

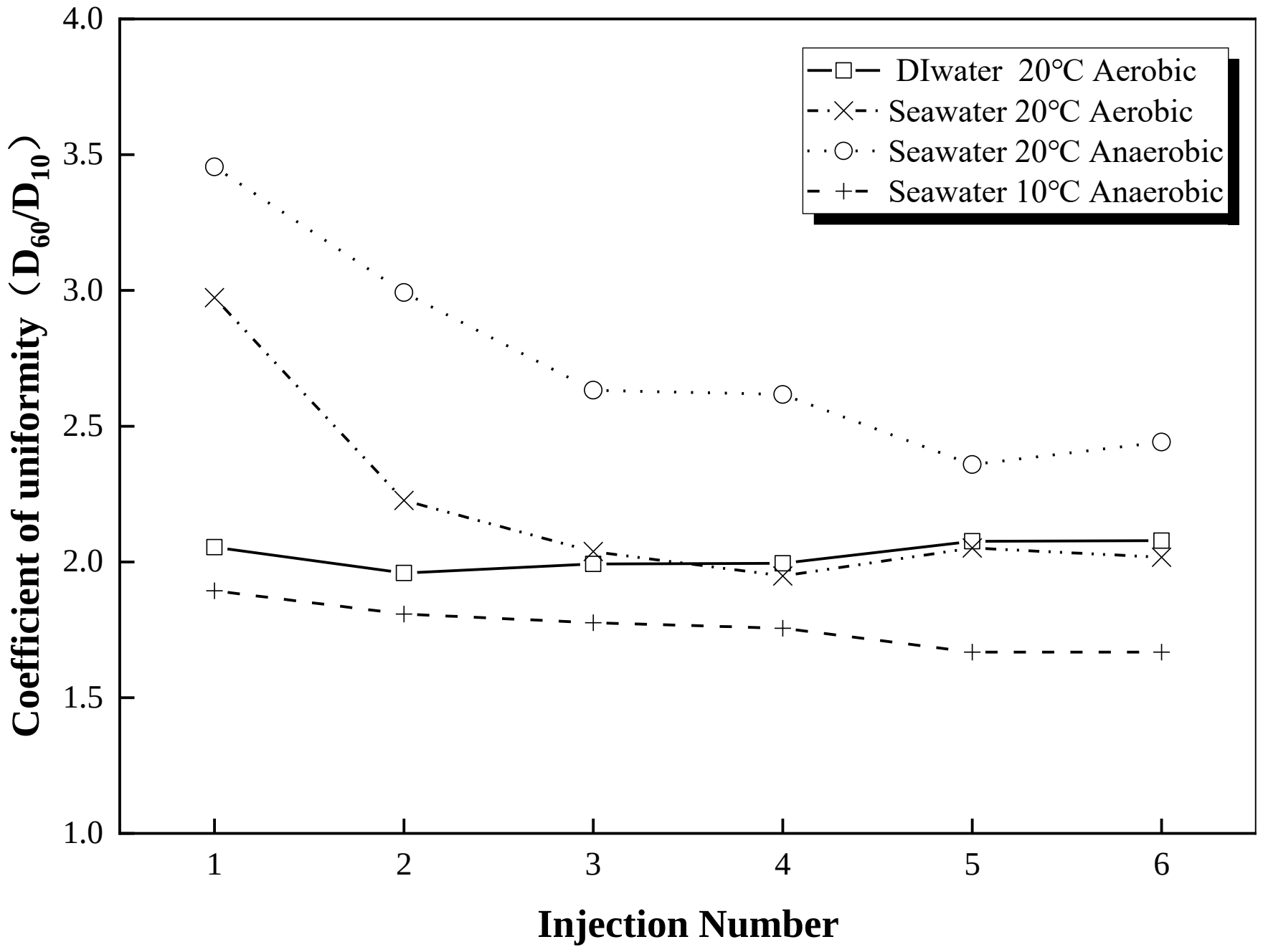

**Fig.13.** The uniformity of calcium carbonate during 1st to 6th cementation solution injection indicated by coefficient of uniformity ($D_{60}/D_{10}$)

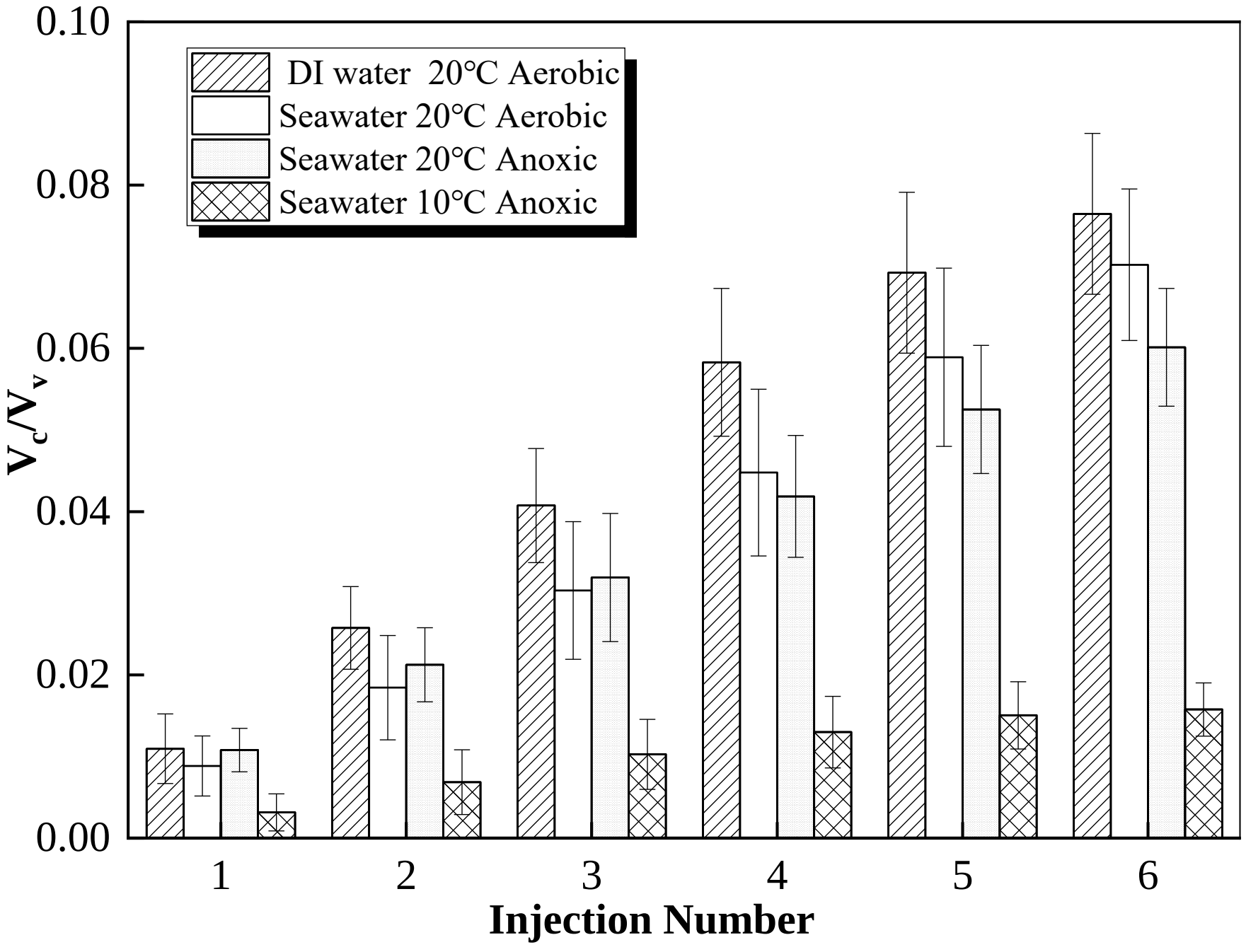

(a)

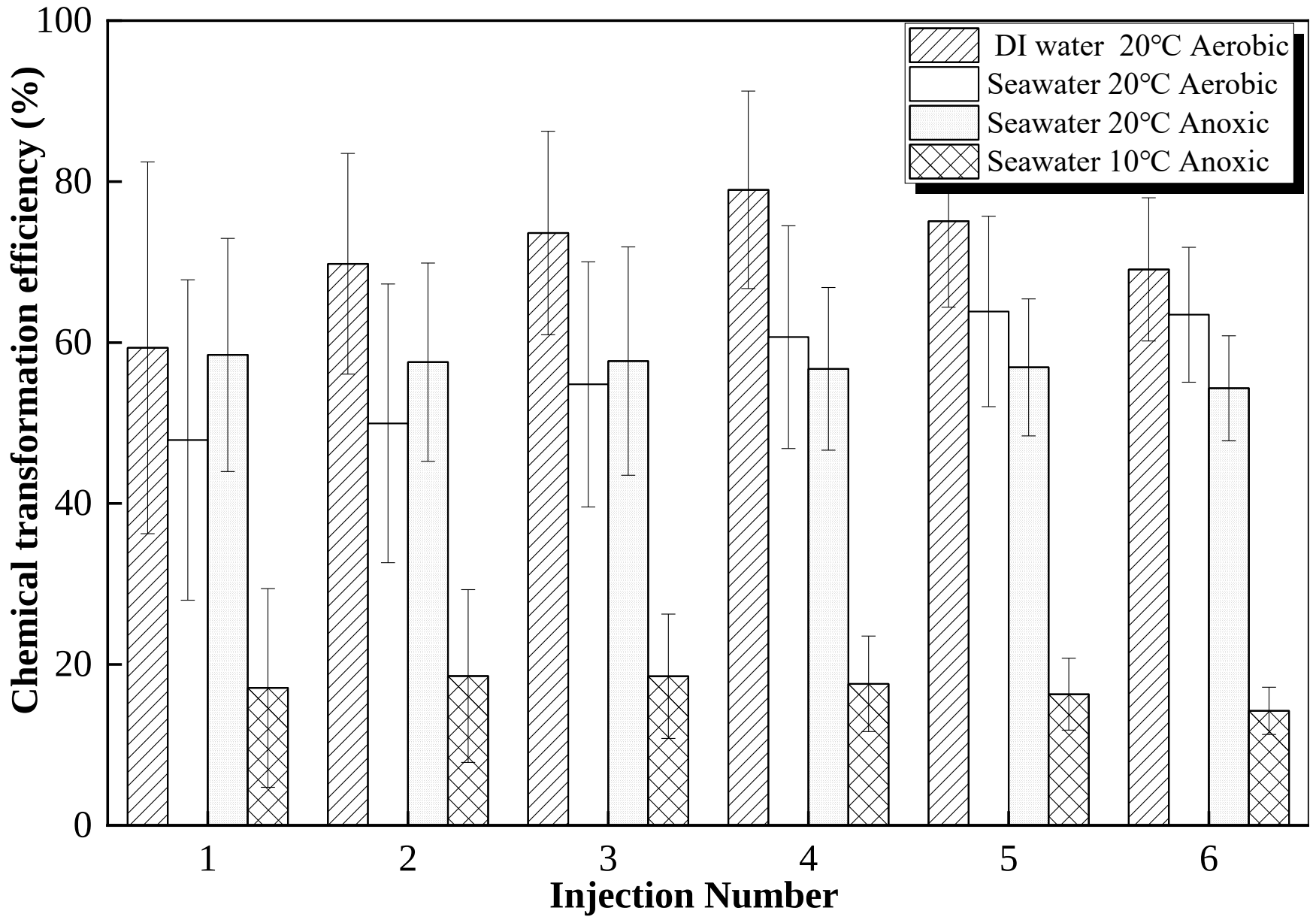

(b)

**Fig.14.** The volume and chemical transformation efficiency of calcium carbonate during 1st to 6th cementation solution injection: (a) the ratio of calcium carbonate volume to pore volume ($V_c/V_v$); (b) chemical transformation efficiency

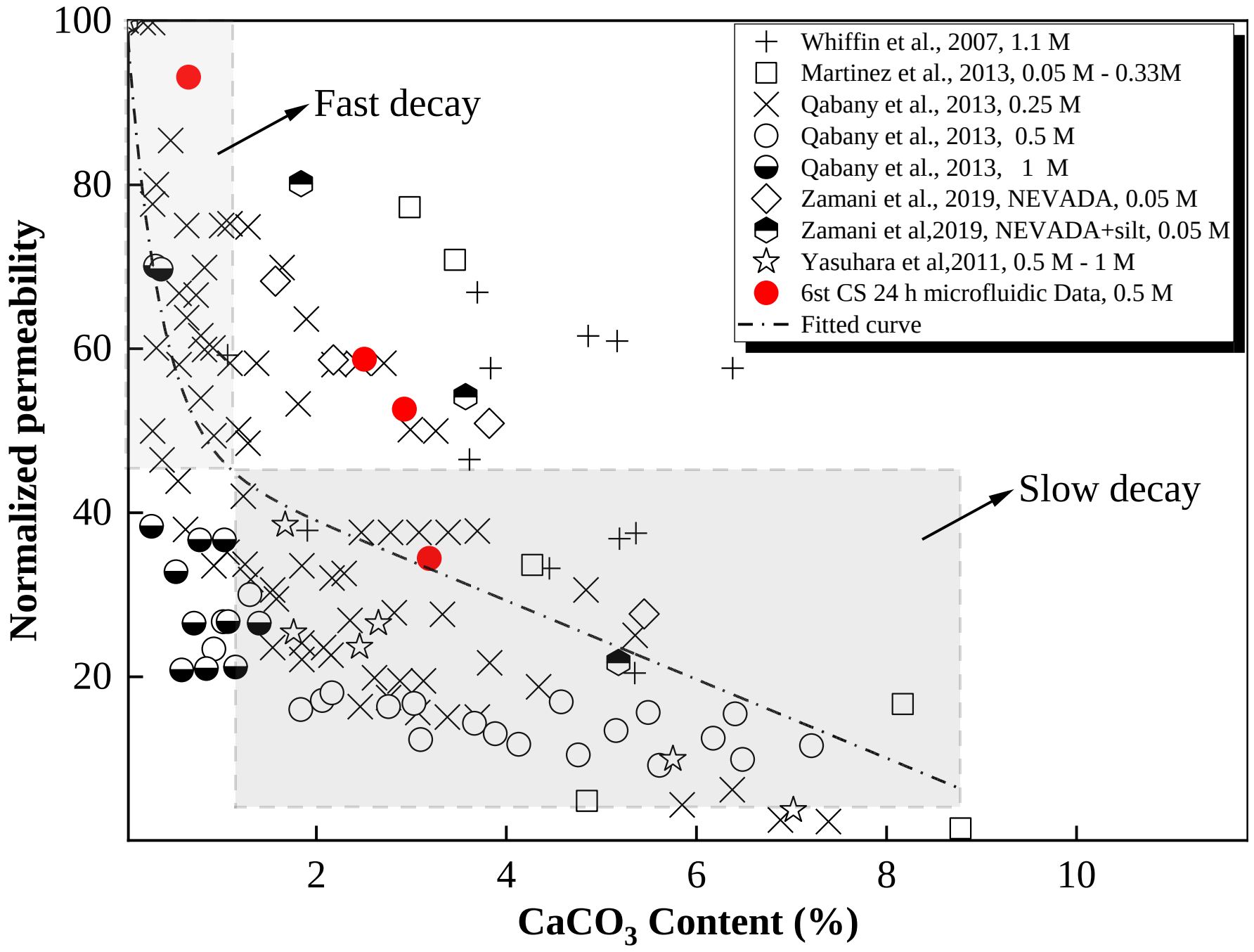

(a)

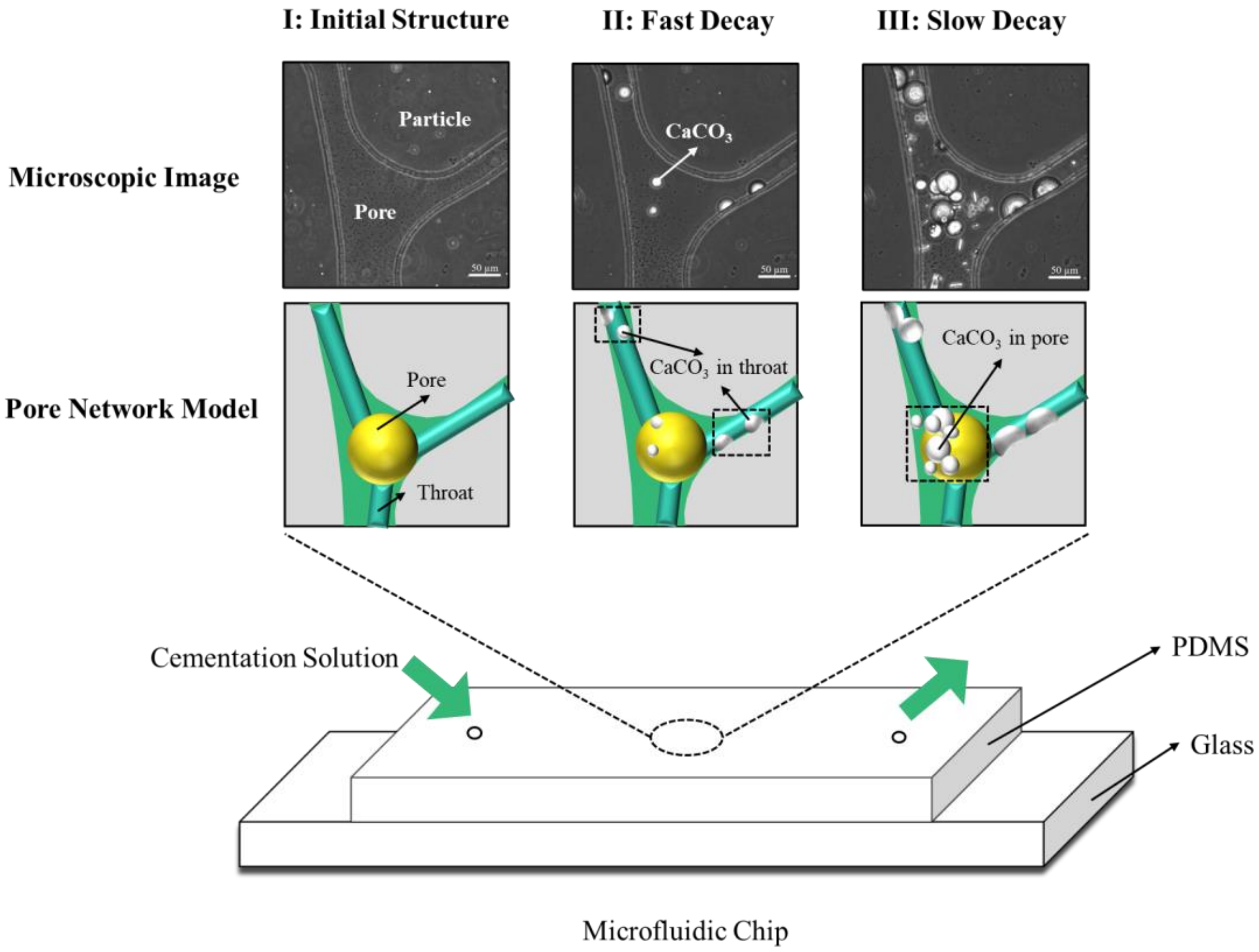

(b)

**Fig.15.** Normalized permeability v.s. $CaCO_3$ content (a) Trend; (b) Mechanism diagram

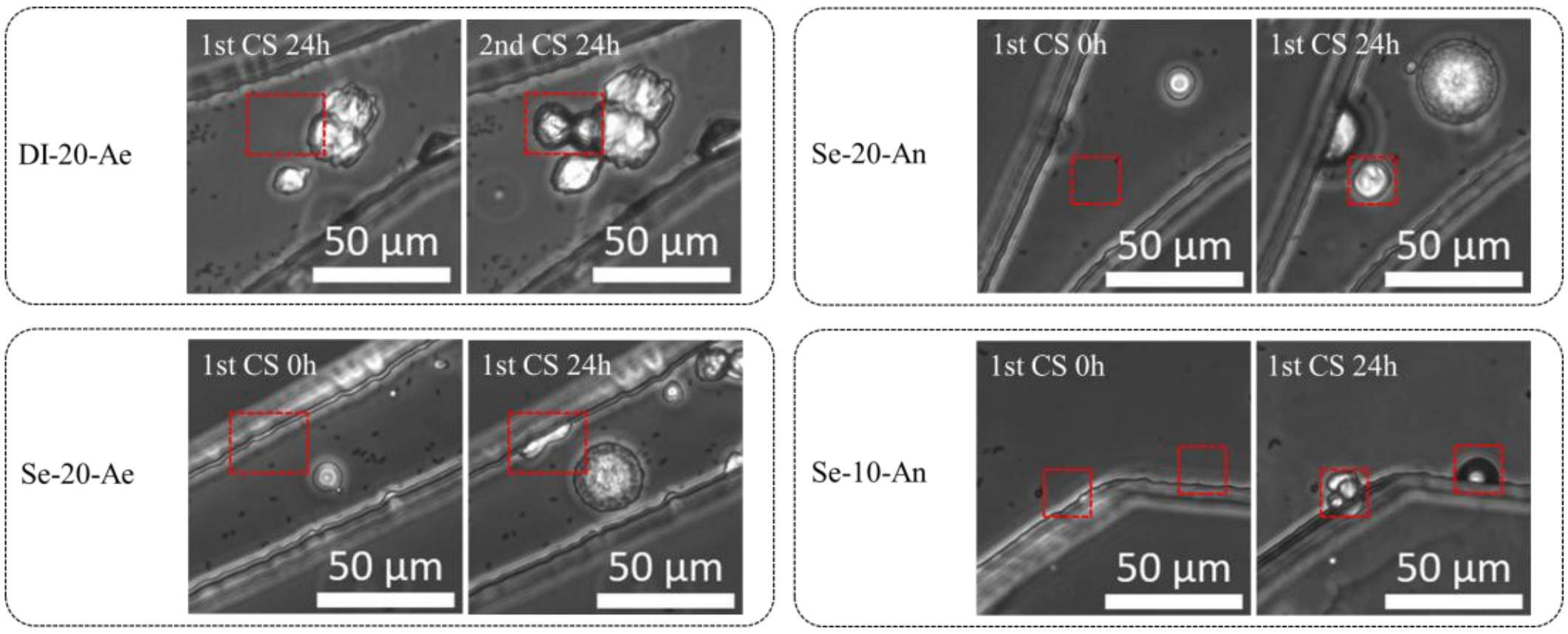

**Fig.16.** Bacteria induce calcium carbonate production